\documentclass[12pt, draftclsnofoot, onecolumn]{IEEEtran}
\usepackage{amssymb}
\usepackage{amsmath}
\usepackage{cite}
\usepackage{url}
\usepackage{xcolor}
\usepackage{cite,graphicx,amsmath,amssymb}
\usepackage{subfigure}
\usepackage{citesort}
\usepackage{fancyhdr}
\usepackage{mdwmath}
\usepackage{mdwtab}
\usepackage{caption}
\usepackage{amsthm}
\usepackage{setspace}
\usepackage{multirow}
\usepackage{epstopdf}
\usepackage{ragged2e}
\usepackage[justification=centering]{caption}
\usepackage{enumerate}
\newtheorem{remark}{Remark}
\newtheorem{theorem}{Theorem}

\newtheorem{lemma}{Lemma}

\newtheorem{corollary}{Corollary}

\usepackage{threeparttable}
\usepackage{algorithm}
\usepackage{algorithmic}
\usepackage{cases}

\renewcommand{\justify}{ \leftskip=0pt \rightskip=0pt plus 0cm}

\begin{document}
	
	\title{Non-Orthogonal Multiple Access For Near-Field Communications }
	\author{
		Jiakuo~Zuo,~\IEEEmembership{Member,~IEEE,}
		Xidong~Mu,~\IEEEmembership{Member,~IEEE,}
		Yuanwei~Liu,~\IEEEmembership{Senior Member,~IEEE,}		
		%        Ertugrul~Basar,~\IEEEmembership{Senior,~IEEE,}
		%        and Octavia A. Dobre,~\IEEEmembership{Fellow,~IEEE}
		\thanks{J. Zuo is with the School of Internet of Things, Nanjing University of Posts and Telecommunications, Nanjing 210003, China (e-mail: zuojiakuo@njupt.edu.cn).}
		\thanks{X. Mu and Y. Liu are with Queen Mary University of London, London E1 4NS, U.K. (email: xidong.mu@qmul.ac.uk, yuanwei.liu@qmul.ac.uk).}
		%\thanks{E. Basar is with the Communications Research and Innovation Laboratory (CoreLab), Department of Electrical and Electronics Engineering, Ko\c{c} University, Sariyer 34450, Istanbul, Turkey (email: ebasar@ku.edu.tr).}
		%\thanks{O. A. Dobre is with the Department of Electrical and Computer Engineering, Memorial University, St. Johns, NL A1C 5S7, Canada (email: odobre@mun.ca).}
	}
	
	%\IEEEspecialpapernotice{(Invited Paper)}
	\maketitle
	%\IEEEspecialpapernotice{(Invited Paper)}
	%\thispagestyle{fancyplain}
	%\pagestyle{fancy}
	\vspace{-2cm}
	\begin{abstract}
		The novel concept of near-field non-orthogonal multiple access (NF-NOMA) communications is proposed. The unique near-filed beamfocusing enables NOMA to be carried out in both \emph{angular} and \emph{distance} domains. Based on the hybrid beamforming transmitter, two novel frameworks are proposed, namely, single-location-beamfocusing NF-NOMA and multiple-location-beamfocusing NF-NOMA. 1) For the single-location-beamfocusing NF-NOMA, two NOMA users in the same angular direction with distinct quality of service (QoS) requirements can be grouped into one cluster. By exploiting the analog beamformers focusing on specific locations, the \emph{far-to-near} successive interference cancellation order and the \emph{distance-domain} user clustering can be further facilitated.
		The hybrid beamformer design and power allocation problem is formulated to maximize the sum rate of the users with higher QoS (H-QoS) requirements, subject to the rate constraints of all users. To solve this problem, the analog beamformer is first designed to focus the energy on the H-QoS users and the zero-forcing (ZF) digital beamformer is further employed to remove the inter-cluster interference. Then, the corresponding optimal power allocation is obtained. 2) For the multiple-location-beamfocusing NF-NOMA, the two NOMA users grouped in the same cluster can have different angular directions and are served by one analog beamformer focusing on multiple locations. To maximize the sum rate of H-QoS users, the analog beamformer is first designed using the beam-splitting technique, which focuses the energy on both two NOMA users at two different locations. Then, a
		singular value decomposition (SVD) based ZF (SVD-ZF) digital beamformer is designed to mitigate the inter-cluster interference. Furthermore, an antenna allocation algorithm is proposed by employing the many-to-one matching method. Finally, an iterative algorithm is proposed to obtain suboptimal power allocation solutions via the fractional programming. Numerical results demonstrate
		that: i) in contrast to the conventional far-field NOMA, the proposed NF-NOMA schemes can achieve a higher spectral efficiency even if the H-QoS users are far located; ii) the multi-user interference can be well mitigated by exploiting beamfocusing in NF-NOMA, and iii) NF-NOMA transmission always outperforms near-field orthogonal multiple access transmission. 
	\end{abstract}
	
	\vspace{-0.5cm}
	\vspace{-0.3cm}
	\section{Introduction}
	The next sixth generation (6G) wireless networks are promising to support immense throughput, ultra-massive communications, and ultra-high spectrum efficiency\textcolor[rgb]{0.00,0.00,0.00}{~\cite{9755276}}. To fulfill these ambitious requirements, extremely large-scale (XL) antenna arrays (hundreds or even thousands of antennas) and high-frequency spectra (millimeter wave (mmWave) and terahertz (THz) bands) {have} to be used. In this context, {the near-field \textcolor[rgb]{0.00,0.00,0.00}{(NF)} wireless propagation in 6G becomes significantly dominated, thus leading to the new paradigm of NF communications (NFC)}~\cite{9903389,Liu2023NearFieldCW}. \textcolor[rgb]{0.00,0.00,0.00}{This is fundamentally different from} previous generations of wireless networks mainly \textcolor[rgb]{0.00,0.00,0.00}{relying} upon far-field communications (FFC), \textcolor[rgb]{0.00,0.00,0.00}{where the electromagnetic wavefronts can be approximated as plane. In the FFC, the radio frequency (RF) signals are {delivered} by antenna arrays via beamsteering~\cite{10068140}, towards a specific direction in the angular domain. By contrast, in the NFC, {the unique spherical wavefronts can be exploited to }generate focused beams in specific spatial region, namely beamfocusing~\cite{Liu2023NearFieldCW}. As a result, NFC provides new \textcolor[rgb]{0.00,0.00,0.00}{{degrees of freedom (DoFs)}} in both angle and distance domains to achieve precise signal enhancement and co-channel interference mitigation~\cite{10068140}.}
	
	\textcolor[rgb]{0.00,0.00,0.00}{Moreover}, to support massive connectivity and enhance spectral efficiency \textcolor[rgb]{0.00,0.00,0.00}{(SE)}, highly efficient next-generation multiple access (NGMA) techniques are vital for 6G {~\cite{9693417}}. As a prominent member of NGMA family, non-orthogonal multiple access (NOMA) provides a higher degree of compatibility and flexibility. {In sharp contrast to orthogonal multiple access (OMA), NOMA allows multiple users to share the same resource block by \textcolor[rgb]{0.00,0.00,0.00}{carrying out the} superposition coding at the transmitter and the successive interference cancellation (SIC) at the receiver~\cite{9693417,Liu2018NonOrthogonalMA}. Therefore, NOMA is capable of achieving higher \textcolor[rgb]{0.00,0.00,0.00}{SE} and supporting massive connectivity. Superiorities of NOMA \textcolor[rgb]{0.00,0.00,0.00}{have} been revealed in \textcolor[rgb]{0.00,0.00,0.00}{multiple-input-multiple-output (MIMO)}  NOMA~\cite{Li2018NOMAAidedCM,Zhang2017CapacityAO,Li2019UnderlaySM}, mmWave NOMA~\cite{Zhu2019MillimeterWaveNW,Dai2018HybridPM,Wang2017SpectrumAE,Wei2018MultiBeamNF,Liu2020MultiBeamNF} and THz NOMA~\cite{10029901,9115278,9605603} communications. However, current research contributions only focused on the NOMA designs in the \textcolor[rgb]{0.00,0.00,0.00}{planar-wave-based far-field (FF) regime}. The great potential of NOMA in NFC has not been unlocked for further performance improvement.}  
	\subsection{Prior Works}
	\subsubsection{Beamfocusing \textcolor[rgb]{0.00,0.00,0.00}{Design} in NFC}
	\textcolor[rgb]{0.00,0.00,0.00}{The NF-beamfocusing provides new opportunities for communication designs and thus attract extensive attention. For example,} in~\cite{Zhang2021BeamFF}, the potential of NF-beamfocusing was studied by considering different antenna structure, \textcolor[rgb]{0.00,0.00,0.00}{namely,} fully-digital architectures, hybrid phase \textcolor[rgb]{0.00,0.00,0.00}{shifter}-based \textcolor[rgb]{0.00,0.00,0.00}{architectures}, and dynamic metasurface antenna architectures. It was shown that the \textcolor[rgb]{0.00,0.00,0.00}{NF-beamfocusing} can \textcolor[rgb]{0.00,0.00,0.00}{further mitigate} co-channel interference in multi-user communication scenarios, which is not achievable by traditional \textcolor[rgb]{0.00,0.00,0.00}{FF-beamsteering}. The \textcolor[rgb]{0.00,0.00,0.00}{NF} physical layer security problem was studied in~\cite{Zhang2023PhysicalLS}. \textcolor[rgb]{0.00,0.00,0.00}{It reveals that} utilizing NF-beamfocusing, the secrecy performance in the NFC systems primarily relies on the relative distance of the eavesdropper with respect to the legitimate user. This observation is {distinct with} the case in FFC systems where the secrecy performance is mainly dependent on the angular disparity between the legitimate user and the eavesdropper with respect to the base station (BS).
	In~\cite{Wang2023NearFieldIS}, a NF integrated sensing and communication (NF-ISAC) framework was proposed and the joint distance and angle estimation of the target was achieved by the framework. Compared with FF-ISAC, {the NF-ISAC can achieve \textcolor[rgb]{0.00,0.00,0.00}{higher accuracy in} the angle and distance estimation. In~\cite{Cui2021NearFieldWB}, an important challenge for NFC, namely the NF beam-split effect, was revealed. To address this challenge, a phase-delay beamfocusing method was proposed to mitigate the NF beam-split effect.} 
	\subsubsection{\textcolor[rgb]{0.00,0.00,0.00}{Hybrid Beamforming Design in FF-NOMA Communications}}
	In the \textcolor[rgb]{0.00,0.00,0.00}{FF-NOMA communications, the users with similar angular directions are grouped as a NOMA cluster and share the same analog beamformer} (the same RF chain). {For example, in~\cite{Zhu2019MillimeterWaveNW}, the users with highly correlated channels were grouped into the same cluster. Whereafter, a joint hybrid beamforming and power allocation optimization problem was formulated to maximize the sum rate.} {The authors of}~\cite{Dai2018HybridPM} applied simultaneous wireless information and power transfer (SWIPT) in {FF-NOMA communications}, and the user grouping, hybrid percoding, power allocation and power splitting were jointly designed to enable the spectrum and energy-efficiency.~{The authors of}~\cite{Wang2017SpectrumAE} proposed a new spectrum and energy efficient transmission scheme, i.e., {beamspace NOMA}, where NOMA was used as multiple access scheme for beamspace MIMO in FFC networks. \textcolor[rgb]{0.00,0.00,0.00}{As a further advance,} {the authors of~\cite{Wei2018MultiBeamNF} proposed a multi-beam NOMA scheme}, where the single analog beamformer was split into multiple sub-analog-beamformers to accommodate the NOMA users with different angular directions. Compared to {single analog beamformer schemes~\cite{Zhu2019MillimeterWaveNW,Dai2018HybridPM,Wang2017SpectrumAE}}, the multi-beam NOMA scheme can provide higher flexibility in serving users and achieve a higher spectral efficiency. 
	Different from~\cite{Wei2018MultiBeamNF}, {a multi-beam beamspace NOMA was proposed in~\cite{Liu2020MultiBeamNF}, where each RF chain can be connected to two or more analog beams. Different from single-beam beamspace NOMA where NOMA transmission are performed among users in the same analog beam, the multi-beam beamspace NOMA enables the users within different analog beams to perform NOMA transmission.}
	\subsection{Motivations and Contributions}
	\textcolor[rgb]{0.00,0.00,0.00}{Despite there are solid research contribution on NOMA designs in FFC, the investigation of NOMA in NFC is still in its fancy. Compared to the conventional FF-NOMA, the primarily benefits brought by NFC for NOMA can be summarized as follows: }
	\begin{itemize}
		\item  \textcolor[rgb]{0.00,0.00,0.00}{Beamfocusing enabled \textit{far-to-near} SIC decoding order}: By exploiting \textcolor[rgb]{0.00,0.00,0.00}{NF-beamfocusing} function, the far-users can have a higher effective channel gain than those near-users. Base on this characteristic, NF-NOMA is able to realize a 'far-to-near' SIC decoding order design. 
		\item \textcolor[rgb]{0.00,0.00,0.00}{Distance-domain NOMA user clustering}:  In NF-NOMA, NF-beamfocusing can split users in the same angular direction into several small clusters, which is generally impossible for FF-NOMA. 
	\end{itemize}
	
	It can be observed that NFC has potential to enhance the flexibility of NOMA transmission. However, to the best of the authors' knowledge, the fundamental practical NF-NOMA transmission framework has not been studied, yet. This provides the main motivation of this work.	
	
	The contributions of this paper can be summarized as follows:
	\begin{itemize}
		\item  \textcolor[rgb]{0.00,0.00,0.00}{We propose the novel concept of NF-NOMA communications, where the NOMA transmission is realized in both angular and distance domains with the aid of the unique NF-beamfocusing property. In particular}, we propose two \textcolor[rgb]{0.00,0.00,0.00}{practical NF-NOMA} frameworks \textcolor[rgb]{0.00,0.00,0.00}{for} the hybrid beamforming transmitter, namely, single-location-beamfocusing NF-NOMA (SLB-NF-NOMA) and multiple-location-beamfocusing NF-NOMA (MLB-NF-NOMA). 
		\item For SLB-NF-NOMA, \textcolor[rgb]{0.00,0.00,0.00}{we group users with high and low QoS (H-/L-QoS) requirements in the same angular direction into one cluster, which is served by an analog beamformer focusing on one specific location. We further distinguish different user clusters in the distance domain. Based on this framework, we maximize the sum rate of H-QoS users by designing the hybrid beamforming and power allocation strategy while satisfying all the users’ QoS requirements.} We propose a three-step \textcolor[rgb]{0.00,0.00,0.00}{algorithm} to solve the formulated problem. In particular, we first propose \textcolor[rgb]{0.00,0.00,0.00}{a single-location-focused analog beamformer} design scheme to focus the energy on the H-QoS users. Then, we employ zero-forcing (ZF) digital beamformer to remove the inter-cluster interference. Finally, we develop an optimal power allocation scheme. 
		\item  \textcolor[rgb]{0.00,0.00,0.00}{For MLB-NF-NOMA, we group H- and L-QoS users distributed at different angular directions into one cluster, which is served by an analog beamformer focusing on multiple locations.} To maximize the sum rate of H-QoS users, \textcolor[rgb]{0.00,0.00,0.00}{we first exploit beam-splitting technique to generate the {multiple-location-focused analog beamformer}.} Then, to manage the inter-cluster interference, we adopt singular value decomposition (SVD) based ZF (SVD-ZF) digital beamformer. After that, we propose a novel antennas allocation algorithm based on many-to-one matching to determine the number of antennas allocated to the H-QoS and L-QoS users. Finally, we propose \textcolor[rgb]{0.00,0.00,0.00}{a fractional programming based iterative algorithm to obtain suboptimal power allocation solutions.}
		\item Our numerical results show that the proposed SLB-NF-NOMA and {MLB-NF-NOMA} schemes achieve a higher sum rate performance than the convectional FF-NOMA based schemes. Furthermore, by exploiting NF-beamfocusing, the proposed {SLB-NF-NOMA} and {MLB-NF-NOMA} schemes is able to significantly mitigate the total interference compared to FF-NOMA based schemes. It is also indicated that the {SLB-NF-NOMA} and {MLB-NF-NOMA} \textcolor[rgb]{0.00,0.00,0.00}{have} superior performance compared \textcolor[rgb]{0.00,0.00,0.00}{to} NF-OMA schemes. 	
	\end{itemize}
	\subsection{Organization and {Notations}}
	The rest of this paper is organized as follows. In section II, a {SLB-NF-NOMA} framework is conceived. Then, the hybrid \textcolor[rgb]{0.00,0.00,0.00}{beamforming} and power allocation schemes are proposed for SLB-NF-NOMA framwork. 
	In section III, {a MLB-NF-NOMA framework is proposed. In addition, the hybrid beamforming, antenna allocation, and power allocation problem are solved separately.} 
	Section IV provides numerical results for characterizing the proposed frameworks and algorithms. Finally, conclusions are drawn in Section V.  
	
	\textbf{\emph{Notations}}: Scalars, vectors, and matrices are denoted by lower-case, boldface lower-case and bold-face upper-case letters, respectively; $\mathbb{C}^{M \times 1}$ denotes the space of 
	$M\times 1$ complex valued vectors. The $(m,n)$-th element of matrix $\textbf{X}$ is denoted as $\left[ \mathbf{X} \right] _{m,n}$. ${\textbf{x}}^{H}$ and ${\textbf{X}}^{H}$ denote the conjugate transpose of vector \textbf{x} and matrix \textbf{X}, respectively. $\mathcal{C}\mathcal{N}\left( 0,\sigma ^2 \right) $ represents the distribution
	of a circularly symmetric complex Gaussian variable (CSCG) with zero mean and $\sigma ^2$ variance.
	%\vspace{-0.3cm}
	
	\section{Single-Location-Beamfocusing NF-NOMA Framework}
	\subsection{System Model}
\begin{figure}[h]
		\centering
		\includegraphics[scale=0.2]{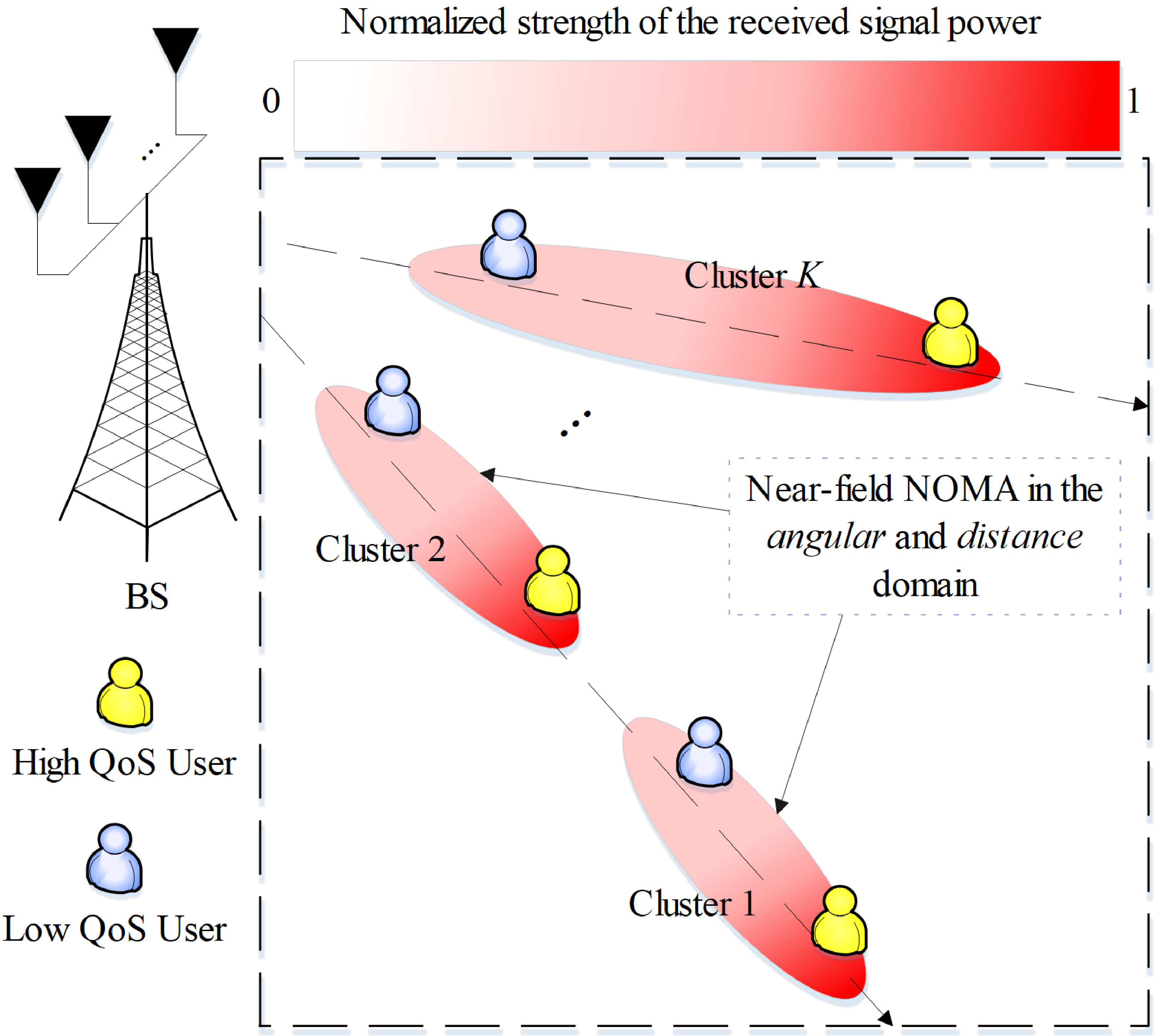}
		\caption{System model of SLB-NF-NOMA}
		\label{Fig1_NF_SB_NOMA}
	\end{figure}

	\textcolor[rgb]{0.00,0.00,0.00}{In this section, we propose a} downlink {SLB-NF-NOMA} communication framework, where \textcolor[rgb]{0.00,0.00,0.00}{a BS employing the hybrid beamformer architecture of} $M_{\mathrm{RF}}$ RF chains and $N_{\mathrm{T}}$ transmit antennas serves $2K$ single-antenna users simultaneously. {An overloaded case is considered, where the number of RF chains is smaller than the number of users, i.e., $M_{\mathrm{RF}}<2K$. For simplicity, we assume that the $2K$ users are grouped into $K$ clusters and each cluster is formed by two users with the same angular direction\textcolor[rgb]{0.00,0.00,0.00}{, as shown in Fig.~\ref{Fig1_NF_SB_NOMA}}. Furthermore, we assume \textcolor[rgb]{0.00,0.00,0.00}{that} the $K$ clusters' data streams in the baseband are precoded by the digital beamforming matrix $\mathbf{W}^{\mathrm{D}}\in \mathbb{C} ^{M_{\mathrm{RF}}\times K}
		$. Then, the digital-domain signal pass the corresponding RF chain and is delivered to $N_{\mathrm{T}}$ phase shifters to perform analog beamforming. Thus, the analog beamformer matrix $\mathbf{W}^{\mathrm{A}}\in \mathbb{C} ^{N_{\mathrm{T}}\times M_{\mathrm{RF}}}$. For achieving higher multiplexing gain,} the number of NOMA clusters is assumed to be equal to the number of RF chains, {i.e.}, $K=M_{\mathrm{RF}}$. \textcolor[rgb]{0.00,0.00,0.00}{Let $\mathcal{M} = \left\{ 1,2,\cdots ,M_{\mathrm{RF}} \right\}$ denote the set of clusters.} Without loss of generality, we assume that the two users in the same cluster have different QoS requirements, namely, one user has a H-QoS requirement and another user has a L-QoS requirement. In each cluster, the NOMA \textcolor[rgb]{0.00,0.00,0.00}{transmission} is applied. Note that specific user pairing strategy \textcolor[rgb]{0.00,0.00,0.00}{can be} employed during the NOMA transmission, which leads to different performance. \textcolor[rgb]{0.00,0.00,0.00}{In this paper, we mainly focus on the} {hybrid beamforming} \textcolor[rgb]{0.00,0.00,0.00}{design} and power allocation after user pairing. 
	
	{Let $\mathcal{U} \left( m,h \right) $ and $\mathcal{U} \left( m,l \right) $ denote H-QoS user and L-QoS user in the \textcolor[rgb]{0.00,0.00,0.00}{$m$}-th cluster, respectively. The NF wireless channel between the BS and user $\mathcal{U} \left( m,k \right) $ is} given by~\cite{Cui2021NearFieldWB,Cui2022NearFieldRW} 
	\begin{equation}\label{hmk}
		\mathbf{g}_{m,k}=\sqrt{N_{\mathrm{T}}}a_{m,k}\mathbf{b}\left( r_{m,k},\theta _{m,k} \right) ,
	\end{equation}
	where $a_{m,k}$, $r_{m,k}$, and $\theta _{m,k}$ denote the free-space path loss, the distance\textbf{,} and the angle-of-departure (AOD) of the user $\mathcal{U} \left( m,k \right) $, respectively. $\mathbf{b}\left( r_{m,k},\theta _{n,k} \right)\in \mathbb{C} ^{M_{\mathrm{RF}}\times 1}$ is the {array response} vector, which can be expressed as 
	\begin{equation}\label{beamfocusing vector}
		\mathbf{b}\left( r_{m,k},\theta _{m,k} \right) =\frac{1}{\sqrt{N_{\mathrm{T}}}}\left[ e^{-j\frac{2\pi}{\lambda}r_{m,k}^{\left( 0 \right)}},e^{-j\frac{2\pi}{\lambda}r_{m,k}^{\left( 1 \right)}},\cdots ,e^{-j\frac{2\pi}{\lambda}r_{m,k}^{\left( N_{\mathrm{T}} \right)}} \right] ^T,
	\end{equation}
	where $r_{m,k}^{\left( n \right)}=\sqrt{\left( r_{m,k}^{2}+\left( \delta _{N_{\mathrm{T}}}^{\left( n \right)}d \right) ^2 \right) -2\delta _{N_{\mathrm{T}}}^{\left( n \right)}dr_{m,k}\sin \theta _{m,k}}$ is the distance between the $n$-th antenna and the user $\mathcal{U} _{m,k}$, $\delta _{N_{\mathrm{T}}}^{\left( n \right)}=n-\frac{N_{\mathrm{T}}-1}{2}$, $n=0,1,\cdots ,N_{\mathrm{T}}-1$
	
	The received signal at user $\mathcal{U} \left( m,k \right) $ is~{given by}
	\begin{equation}\label{signal_ymk}
		y_{m,k}=\underset{\mathrm{desired}~\mathrm{signal}}{\underbrace{g_{m,k}\sqrt{p_{m,k}}s_{mk}}}+\underset{\mathrm{intra}-\mathrm{cluster}~\mathrm{interference}}{\underbrace{g_{m,k}\sqrt{p_{m,\widetilde{k}}}s_{m,\widetilde{k}}}}+\underset{\mathrm{inter}-\mathrm{cluster}~\mathrm{interference}}{\underbrace{\sum_{i\ne m,i\in \mathcal{M}}{g_{i,\left( m,k \right)}\left( \sum_{j\in \left\{ l,h \right\}}{\sqrt{p_{i,j}}s_{i,j}} \right)}}}+\underset{\mathrm{noise}}{\underbrace{z_{m,k}}},
	\end{equation}
	where $g_{m,k}=\mathbf{g}_{m,k}^{H}\mathbf{W}^{\mathrm{A}}\mathbf{w}_{m}^{\mathrm{D}}
	$ {denotes} the effective channel, {$\mathbf{w}_{m}^{\mathrm{D}}\in \mathbb{C} ^{M_{\mathrm{RF}}\times 1}$ is the $m$-th column of digital beamformer matrix $\mathbf{W}^{\mathrm{D}}$}, $g_{i,\left( m,k \right)}=\mathbf{g}_{m,k}^{H}\mathbf{W}^{\mathrm{A}}\mathbf{w}_{i}^{\mathrm{D}}
	$ denotes the interference channel, and $z_{m,k}\sim \mathcal{C} \mathcal{N} \left( 0,\sigma ^2 \right) $ is the additive white Gaussian noise.
	
	%\subsection{Decoding Order and SIC Condition}
	\textcolor[rgb]{0.00,0.00,0.00}{In each cluster}, we assume that \textcolor[rgb]{0.00,0.00,0.00}{H-QoS user} $\mathcal{U} \left( m,h \right) $ first decodes the signal of \textcolor[rgb]{0.00,0.00,0.00}{the L-QoS user} $\mathcal{U} \left( m,l \right) $ and then
	subtracts it from its observation to decode its own information. 
	Therefore, the achievable rate for user $\mathcal{U} \left( m,h \right) $ to decode user $\mathcal{U} \left( m,l \right) $'s signal is given by
	\begin{equation}\label{gamma_21}
		R_{m,l\rightarrow h}=\log _2\left( 1+\frac{p_{m,l}\left| g_{m,h} \right|^2}{p_{m,h}\left| g_{m,h} \right|^2+I_{m,h}^{\mathrm{inter}}\left( \mathbf{p}_{-m} \right) +\sigma ^2} \right),  
	\end{equation}
	where $I_{m,h}^{\mathrm{inter}}\left( \mathbf{p}_{-m} \right) =\sum_{i\ne m,i\in \mathcal{M}}{P_i\left| g_{i,\left( m,h \right)} \right|^2}
	$ is the inter-cluster interference at user $\mathcal{U} \left( m,h \right) $, $P_i=\sum_{k\in \left\{ h,l \right\}}{p_{i,k}}
	$ is the transmit power allocated to cluster $i$, $\mathbf{p}=\left[ p_{1,h},p_{1,l},\cdots ,p_{M_{\mathrm{RF}},h},p_{M_{\mathrm{RF}},l} \right] ^T$, and $\mathbf{p}_{-m}$ is the vector of $\mathbf{p}$ removing the elements $p_{m,l}$ and $p_{m,h}$.
	
	If \textcolor[rgb]{0.00,0.00,0.00}{SIC} is successful, user $\mathcal{U} \left( m,l \right) $ then removes the signal of user $\mathcal{U} \left( m,h \right) $ and decodes \textcolor[rgb]{0.00,0.00,0.00}{its intended signal}. Based on~\eqref{signal_ymk}, the achievable rate for user $\mathcal{U} \left( m,h \right) $ to decode its own
	message is given by
	\begin{equation}\label{gamma_11}
		R_{m,h}=\log _2\left( 1+\frac{p_{m,h}\left| g_{m,h} \right|^2}{I_{m,h}^{\mathrm{inter}}\left( \mathbf{p}_{-m} \right) +\sigma ^2} \right). 
	\end{equation}
	
	\textcolor[rgb]{0.00,0.00,0.00}{For} $\mathcal{U} \left( m, l\right) $, it directly decodes its own signal by treating the other users’ signals as interference. Based on~\eqref{signal_ymk},
	the achievable rate for user $\mathcal{U} \left( m, l\right) $ to decode its own signal is given by
	\begin{equation}\label{gamma_22}
		R_{m,l\longrightarrow l}=\log _2\left( 1+\frac{p_{m,l}\left| g_{m,l} \right|^2}{p_{m,h}\left| g_{m,l} \right|^2+I_{m,l}^{\mathrm{inter}}\left(\mathbf{p}_{-m} \right) +\sigma ^2} \right), 
	\end{equation}
	where  $I_{m,l}^{\mathrm{inter}}\left( \mathbf{p}_{-m} \right) =\sum_{i\ne m,i\in \mathcal{M}}{P_i\left| g_{i,\left( m,l \right)} \right|^2}
	$ is the inter-cluster interference at user $\mathcal{U} \left( m,l \right) $.
	
	To guarantee that the SIC is also performed successively, the SIC condition\textcolor[rgb]{0.00,0.00,0.00}{~\cite{Liu2020MultiBeamNF,Cui2018UnsupervisedML}}, i.e., $\gamma _{m,l\rightarrow h}\geqslant \gamma _{m,l}$, should be satisfied, which can be further expressed as
	\begin{equation}\label{SIC_constaint}
		\frac{\left| g_{m,h} \right|^2}{I_{m,h}^{\mathrm{inter}}\left( \mathbf{p}_{-m} \right) +\sigma ^2}\geqslant \frac{\left| g_{m,l} \right|^2}{I_{m,l}^{\mathrm{inter}}\left( \mathbf{p}_{-m} \right) +\sigma ^2}.
	\end{equation}
	
	Note that the SIC condition~\eqref{SIC_constaint} is also relevant to the power allocation coefficients. If ~\eqref{SIC_constaint} can not be guaranteed, the achievable rate of user $\mathcal{U} \left( m, l\right) $ is \textcolor[rgb]{0.00,0.00,0.00}{degraded into} $R_{m,l}=R_{m,l\longrightarrow h}$. Finally, the achievable rate of user $\mathcal{U} \left( m, l\right) $ can be expressed as~\cite{9729087}
	\begin{equation}\label{RLL}
		\setlength{\abovedisplayskip}{5pt}
		\setlength{\belowdisplayskip}{5pt}
		R_{m,l}=\min \left\{ R_{m,l\longrightarrow h},R_{m,l\longrightarrow l} \right\}, 
	\end{equation} 
	where $m\in \mathcal{M}$.	
	
	\begin{remark}
		The main benefits of the proposed SLB-NF-NOMA framework can be summarized as follows. Firstly, 'far-to-near' SIC decoding order among NOMA users can be realized. \textcolor[rgb]{0.00,0.00,0.00}{As illustrated} in Fig.~\ref{Fig1_NF_SB_NOMA}, with the aid of NF beamfocusing, even if the H-QoS users located far from the BS can achieve a higher effective channel gain than those L-QoS users located near the BS. \textcolor[rgb]{0.00,0.00,0.00}{This contributes to the satisfaction of SIC condition~\eqref{SIC_constaint}, thus guaranteeing the performance of H-QoS users. This is in generally impossible to achieve by FF-NOMA, where users in the same angular direction have a descending channel gain with respect to the distance.}
		Secondly, by exploiting the additional distance domain in NFC, the proposed SLB-NF-NOMA can further group the users in the same angular direction into different small clusters in distance-domain. For example in Fig.~\ref{Fig1_NF_SB_NOMA}, \textcolor[rgb]{0.00,0.00,0.00}{the four users in cluster 1 and cluster 2 are located in the same angular direction and can be splitted into two small clusters, i.e., cluster 1 and cluster 2, in distance-domain. However, for FF-NOMA, the four users have to be grouped into the same cluster.} \textcolor[rgb]{0.00,0.00,0.00}{The advantages of the proposed SLB-NF-NOMA will be numerically demonstrated in Section IV.} 
	\end{remark}  
	\subsection{Problem Formulation }
	{In this context, our aim is to maximize the sum rate of H-QoS users, subject to the constraints on both the minimum QoS requirements of the \textcolor[rgb]{0.00,0.00,0.00}{H- and L-QoS users.} Accordingly, the optimization problem in this paper is formulated as follows:}
	\begin{subequations}\label{OP_singleBF}
		\begin{align}
			&\underset{\mathbf{w}_{m}^{\mathrm{A}},\mathbf{w}_{m}^{\mathrm{D}},p_{m,k}>0}{\max}\sum_{m\in \mathcal{M}}{R_{m,h}} \\
			&\mathrm{s}.\mathrm{t}.~R_{m,k}\geqslant R_{m,k}^{\min},\label{OP_singleBF:b} \\ 
			&   \ \ \ \ \ \sum_{m\in \mathcal{M}}{\sum_{k\in \left\{ h,l \right\}}{p_{m,k}}}\leqslant P_{\max},
			\label{OP_singleBF:c} \\
			&   \ \ \ \ \ \left| \left[ \mathbf{w}_{m}^{\mathrm{A}} \right] _n \right|=\frac{1}{\sqrt{N_{\mathrm{T}}}}, \label{OP_singleBF:d} \\
			&   \ \ \ \ \ \left\| \mathbf{w}_{m}^{\mathrm{D}} \right\| _2=1, \label{OP_singleBF:e}		     
		\end{align} 
	\end{subequations}
	where $\mathbf{w}_{m}^{\mathrm{A}}\in \mathbb{C} ^{N_{\mathrm{T}}\times 1}
	$ is the $m$-th column of analog beamformer matrix $\mathbf{W}^{\mathrm{A}}$, $P_{\max}$ is the maximum transmit power at the BS, $R_{m,k}^{\min}$ is the minimum QoS requirement for user $\mathcal{U} \left( m,k \right) $, $k\in \left\{ h,l \right\} $, $m\in \mathcal{M}$. {Constraint~\eqref{OP_singleBF:b} \textcolor[rgb]{0.00,0.00,0.00}{represents} the minimum QoS requirement of each NOMA user, constraint~\eqref{OP_singleBF:c}
		limits the maximum transmit power at the BS, constraint~\eqref{OP_singleBF:d}
		\textcolor[rgb]{0.00,0.00,0.00}{represents} the constant modulus constraint of analog beamformer, and constraint~\eqref{OP_singleBF:e} is the normalized digital beamformer constraint.}
	
	{\textcolor[rgb]{0.00,0.00,0.00}{It can be observed that} problem~\eqref{OP_singleBF} is a non-convex
		optimization problem due to the non-convex objective function
		and the non-convex minimum QoS requirement constraint~\eqref{OP_singleBF:b}, where the
		transmit power $\left\{ p_{m,k} \right\} 
		$, the analog beamformer $\left\{ \mathbf{w}_{m}^{\mathrm{A}} \right\} 
		$, and the digital beamformer $\left\{ \mathbf{w}_{m}^{\mathrm{D}} \right\} 
		$, are highly coupled. It is non-trivial to find the globally optimal solution. In the following, we propose a three-step algorithm to find a high-quality suboptimal solution.}
	\subsection{Problem Solution}	
	{In this section, we develop a three-step algorithm to solve the original problem~\eqref{OP_singleBF}. Specifically, we first design the optimal analog beamformer to obtain the maximum antenna array gain according to the H-QoS user's location. Then, a ZF digital beamformer is employed for canceling inter-cluster interference. After that, optimal power allocation is performed to maximize the sum rate of H-QoS users. }
	\subsubsection{{Single-Location-Focused Analog Beamformer Design}} 
	We invoke the optimal NF analog beamformer which means that the BS tries to focus the {energy} on the desired locations. As a result, high beamfocusing gains can be obtained. In our considered system, the optimal NF analog beamformer can be obtained by maximizing the antenna array gain for the H-QoS users. More particularly, the \textcolor[rgb]{0.00,0.00,0.00}{proposed single-location--focused analog beamformer} $\mathbf{w}_{m}^{\mathrm{A}}$ can be obtained by maximizing $\left| \mathbf{b}_{m,h}^{H}\left( r_{m,h},\theta _{m,h} \right) \mathbf{w}_{m}^{\mathrm{A}} \right|^2$, where $m\in \mathcal{M}$. Thus, we have
	\begin{equation}\label{single_beamfocusing}
		\left[ \mathbf{w}_{m}^{\mathrm{A}} \right] _n=\frac{1}{\sqrt{N_{\mathrm{T}}}}\mathrm{arg}\left( \left[ \mathbf{b}_{m,h} \right] _n \right) =\frac{1}{\sqrt{N_{\mathrm{T}}}}e^{-j\frac{2\pi}{\lambda}r_{m,h}^{\left( n \right)}}.
	\end{equation}
	\begin{remark}
		~\eqref{single_beamfocusing} implies that the optimal analog beamformer for NFC system should align both the spatial angle $\theta _{m,h}
		$ and BS-user distance $r_{m,h}$
		, which {is} significantly different from analog beamformer design for conventional FFC systems. In addition, if $r_{m,h}
		$ is sufficiently large, the NF {array response} vector $\mathbf{b}\left( r_{m,k},\theta _{m,k} \right) 
		$ in~\eqref{beamfocusing vector} \textcolor[rgb]{0.00,0.00,0.00}{is degraded into the conventional FF {array response} vector $\widetilde{\mathbf{b}}\left(\theta _{m,k} \right) =\frac{1}{\sqrt{N_{\mathrm{T}}}}\left[ 1,e^{-j\pi \theta _{m,k}},\cdots ,e^{-j\pi \left( N_{\mathrm{T}}-1 \right) \theta _{m,k}} \right] ^T
			$.	}
	\end{remark}
	
	Based on the above \textcolor[rgb]{0.00,0.00,0.00}{single-location-focused analog beamformer designed in}~\eqref{single_beamfocusing}, we have: $\left| \mathbf{b}_{m,h}^{H}\left( r_{m,h},\theta _{m,h} \right) \mathbf{w}_{m}^{\mathrm{A}} \right|^2 = 1$, which indicates that the antenna array gain at user $\mathcal{U} \left( m, h\right) $ achieves its maximum value. For an arbitrary user location $\left( r,\theta \right)$, the normalized antenna array gain is as follows:
	\begin{equation}\label{AAgain_single}
		\left| \mathbf{b}^H\left( r,\theta \right) \mathbf{w}_{m}^{\mathrm{A}} \right|=\frac{1}{N_{\mathrm{T}}}\left| \sum_{n=0}^{N_{\mathrm{T}}-1}{e^{j\frac{2\pi}{\lambda}\left( r^{\left( n \right)}-r_{m,h}^{\left( n \right)} \right)}} \right|=A^{\mathrm {SA}}_{\mathrm{gain}}\left( r,\theta ,r_{m,h},\theta _{m,h} \right) .
	\end{equation}
	
\begin{figure}[t]
		\centering
		\includegraphics[scale=0.6]{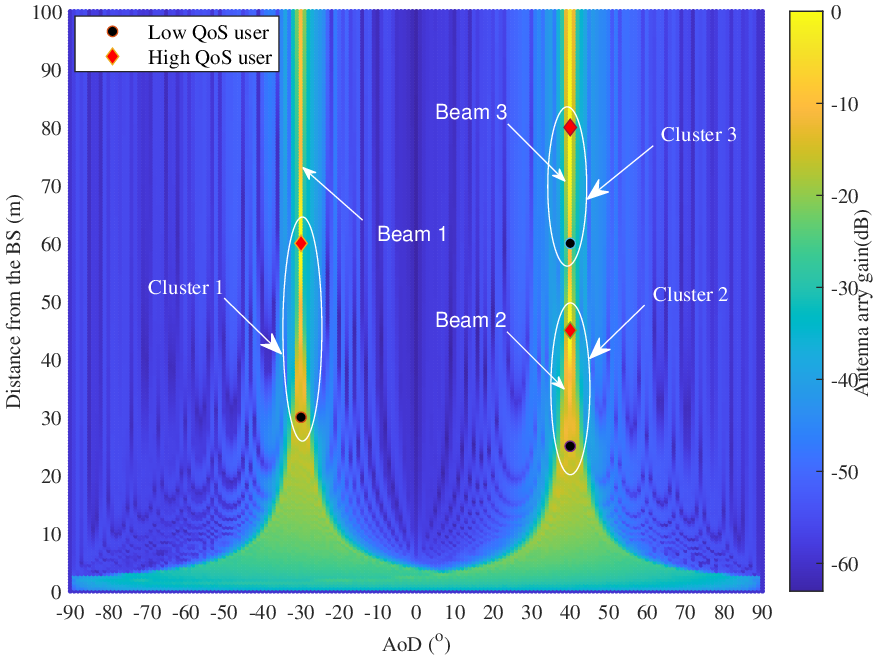}
		\caption{\justify{\normalsize {Antenna array gain map for SLB-NF-NOMA system. We assume $N_{\mathrm{T}}=1024$, the locations $\left( r_{m,h},\theta _{m,h} \right)$ and $\left( r_{m,L},\theta _{m,L} \right)$ of H-QoS and L-QoS users in cluster 1, cluster 2, and cluster 3 are $\left( 30\mathrm{m},-30^{\mathrm{o}} \right) $ and $\left( 60\mathrm{m},-30^{\mathrm{o}} \right) $, $\left( 25\mathrm{m},40^{\mathrm{o}} \right) $ and $\left( 45\mathrm{m},40^{\mathrm{o}} \right) $, $\left( 60\mathrm{m},40^{\mathrm{o}} \right)$ and $\left( 80\mathrm{m},40^{\mathrm{o}} \right)$, respectively.}}}
		\label{single_Beam}
	\end{figure}

	It {can be observed} that the maximum value of $A^{\mathrm {SA}}_{\mathrm{gain}}\left( r,\theta ,r_{m,h},\theta _{m,h} \right) 
	$ is 1 and the array gain reaches its peak when $r=r_{m,h}$ and $\theta =\theta _{m,h}
	$, which means that the beam energy is exactly focused on the location $\left( r_{m,h},\theta _{m,h} \right) 
	$. {An example of} the antenna array gain map for SLB-NF-NOMA system is shown in Fig.~\ref{single_Beam} to illustrate \textcolor[rgb]{0.00,0.00,0.00}{the single-location--focused analog beamformer. We can observe that the generated analog beamformer for each cluster is able to exactingly focus the beam energy around the desired H-QoS users' locations. As a result, the far H-QoS users can achieve a higher effective channel gain than the near L-QoS users, resulting in \textit{far-to-near} SIC decoding order. In addition,  even if there are four users in the same angular direction, the NF-NOMA can distinguish different user clusters in the distance domain and generate two beams to serve the NOMA users in each cluster.}
	\subsubsection{ZF Digital {Beamformer} Design}
	After obtaining the analog beamformer $\left\{ \mathbf{w}_{m}^{\mathrm{A}} \right\} 
	$, the equivalent channel vector for user $\mathcal{U} \left( m,k \right) $ can be defined as $\widetilde{\mathbf{g}}_{m,k}^{H}=\mathbf{g}_{m,k}^{H}\mathbf{W}^{\mathrm{A}}$. Since the beamspace channel vectors of the H- and L-QoS users in the same cluster are highly correlated, we can use the beamspace channel vector of the H-QoS user as the cluster's equivalent channel vector. Therefore, we adopt \textcolor[rgb]{0.00,0.00,0.00}{the} ZF digital beamformer design~\cite{Wang2017SpectrumAE} to remove the H-QoS users' inter-cluster interference, which means that $\left\{ \begin{array}{c}
		\widetilde{\mathbf{g}}_{i,h}^{H}\mathbf{w}_{m}^{\mathrm{D}}=0,i\ne m\\
		\widetilde{\mathbf{g}}_{i,h}^{H}\mathbf{w}_{m}^{\mathrm{D}}\ne 0,i=m\\
	\end{array} \right. $. It should be noted that the main reason to design the digital beamformer with respect to the H-QoS users is that the H-QoS user \textcolor[rgb]{0.00,0.00,0.00}{have to} decode the L-QoS user's signal before its own signal. Thus, the achievable rate of the H-QoS user is not affected by other users' signal. Let $\widetilde{\mathbf{G}}=\left[ \widetilde{\mathbf{g}}_{1,h},\widetilde{\mathbf{g}}_{2,h},\cdots ,\widetilde{\mathbf{g}}_{M_{\mathrm{RF}},h} \right] $ be the equivalent channel matrix for all clusters. Then, the digital beamformer matrix $\widetilde{\mathbf{W}}^{\mathrm{D}}$ can be expressed as
	\begin{equation}\label{ZF_DBM}
		\widetilde{\mathbf{W}}^{\mathrm{D}}=\left[ \widetilde{\mathbf{w}}_{1}^{\mathrm{D}},\widetilde{\mathbf{w}}_{2}^{\mathrm{D}},\cdots ,\widetilde{\mathbf{w}}_{M_{\mathrm{RF}}}^{\mathrm{D}} \right] =\widetilde{\mathbf{G}}\left( \widetilde{\mathbf{G}}^H\widetilde{\mathbf{G}} \right) ^{-1}.  
	\end{equation}
	
	By normalizing the vectors $\left\{ \widetilde{\mathbf{w}}_{m}^{\mathrm{D}} \right\} $, the digital beamformer for the $m$-th cluster can be expressed as 
	\begin{equation}\label{ZF_DB}
		\mathbf{w}_{m}^{\mathrm{D}}=\frac{\widetilde{\mathbf{w}}_{m}^{\mathrm{D}}}{\left\| \widetilde{\mathbf{w}}_{m}^{\mathrm{D}} \right\| _2}.
	\end{equation} 
	
	With the obtained digital beamformer ~\eqref{ZF_DB}, we have $I_{m,h}^{\mathrm{inter}}\left( \mathbf{p}_{-m} \right) =\sum_{i\ne m,i\in \mathcal{M}}{P_i\left| \widetilde{\mathbf{g}}_{m,h}^{H}\mathbf{w}_{i}^{\mathrm{D}} \right|^2}=0$, which indicates that the inter-cluster interference is perfectly eliminated on the H-QoS user in each cluster. As a result, the achievable rate in~\eqref{gamma_21} and~\eqref{gamma_11} can be rewritten respectively as
	\begin{equation}\label{ZF_gamma_21}
		R_{m,l\rightarrow h}=\log _2\left( 1+\frac{p_{m,l}\left| g_{m,h} \right|^2}{p_{m,h}\left| g_{m,h} \right|^2+\sigma ^2} \right). 
	\end{equation}
	\begin{equation}\label{ZF_gamma_11}
		\setlength{\abovedisplayskip}{5pt}
		\setlength{\belowdisplayskip}{5pt}
		R_{m,h}=\log _2\left( 1+\frac{p_{m,h}\left| g_{m,h} \right|^2}{\sigma ^2} \right). 
	\end{equation}
	
	\subsubsection{Power Allocation Optimization}
	For given analog beamformer $\left\{ \mathbf{w}_{m}^{\mathrm{A}} \right\} 
	$ and digital beamformer $\left\{ \mathbf{w}_{m}^{\mathrm{D}} \right\} 
	$, the power allocation optimization problem can be written as follows:
	\begin{subequations}\label{SB_power_allocation}
		\begin{align}
			&\underset{p_{m,h},p_{m,l}>0}{\max}\sum_{m\in \mathcal{M}}{R_{m,h}} \\
			&\mathrm{s}.\mathrm{t}.~p_{m,h}\geqslant \frac{r_{m,k}^{\min}\sigma ^2}{\left| g_{m,h} \right|^2}
			,\label{OP_singleBF_rewritten:b} \\ 
			&   \ \ \ \ \ p_{m,l}\geqslant r_{m,k}^{\min}\left( p_{m,h}+\frac{\sigma ^2}{\left| g_{m,h} \right|^2} \right) 
			, \label{OP_singleBF_rewritten:c} \\
			&   \ \ \ \ \ 
			p_{m,l}\geqslant r_{m,k}^{\min}\left( p_{m,h}+\frac{I_{m,l}^{\mathrm{inter}}\left( \mathbf{p}_{-m} \right) +\sigma ^2}{\left| g_{m,l} \right|^2} \right), \\
			&   \ \ \ \ \ 	\eqref{OP_singleBF:c}, 	    
		\end{align} 
	\end{subequations}
	where $m\in \mathcal{M}$.
	
	It is easy to observe that problem~\eqref{SB_power_allocation} is a convex optimization problem, which be efficiently solved with convex optimization software, such as CVX~\cite{cvx}.
	\subsubsection{Overall Algorithm and Complexity Analysis}
	The overall algorithm of hybrid beamforming and power allocation for SLB-NF-NOMA system is sketched in \textbf{Algorithm~\ref{Single-beamfocusing}}. It is observed that the complexity of \textbf{Algorithm~\ref{Single-beamfocusing}} mainly depends on that of solving problem~\eqref{SB_power_allocation}. By utilizing numerical convex program solves, e.g., interior-point method, the computational complexity of solving problem is $\mathcal{O} \left( \left( 2M_{\mathrm{RF}} \right) ^{3.5} \right) 
	$~\cite{Hao2019CodebookBasedME}, where $2M_{\mathrm{RF}}$ denotes the number of \textcolor[rgb]{0.00,0.00,0.00}{optimization} variables.
	\begin{algorithm}
		\caption{The proposed scheme for SLB-NF-NOMA system}
		\label{Single-beamfocusing}
		\begin{algorithmic}[1]
			\STATE  \textbf{Single-Location-Focused Analog Beamformer Design}
			\STATE ~~The analog beamformer $\mathbf{w}_{m}^{\mathrm{A}}$ is calculated via~\eqref{single_beamfocusing}, $m\in \mathcal{M}$;
			\STATE  \textbf{Digital Beamformer Design}
			\STATE  ~~Calculate the digital beamformer $\mathbf{w}_{m}^{\mathrm{D}}$ according to~\eqref{ZF_DB}, $m\in \mathcal{M}$;
			\STATE  \textbf{Power Allocation} 
			\STATE   ~~Obtain power allocation strategy $\mathbf{p}$ by solving convex optimization problem\textcolor[rgb]{0.00,0.00,0.00}{~\eqref{SB_power_allocation}};
		\end{algorithmic}
	\end{algorithm} 
	%\vspace{-1cm}
	\section{Multiple-Location-Beamfocusing NF-NOMA Framework}
	\subsection{System Model}
\begin{figure}[h]
		\setlength{\belowcaptionskip}{-20pt}
		\centering
		\includegraphics[scale=0.3]{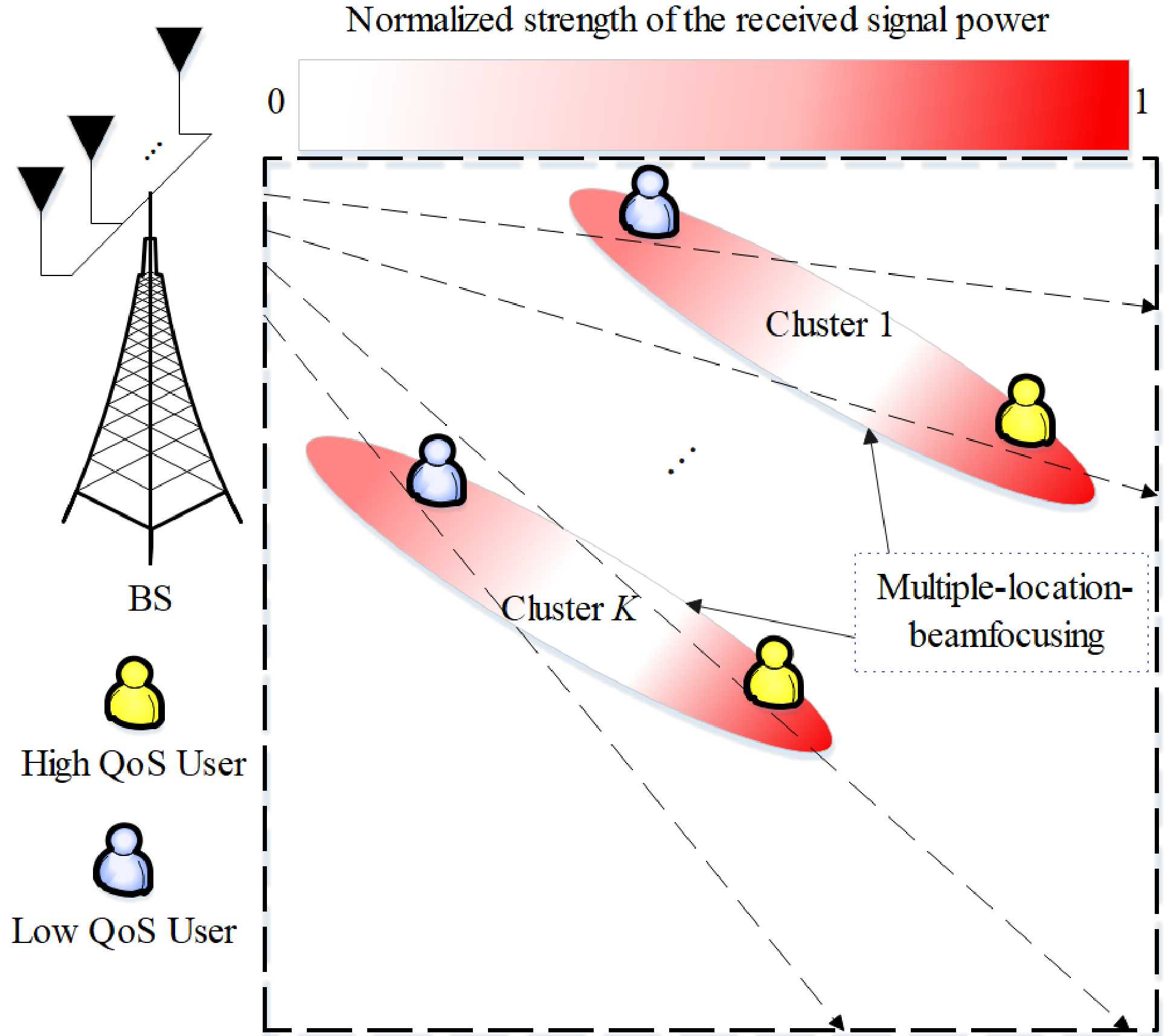}
		\caption{System model of MLB-NF-NOMA}
		\label{NF_MABC_NOMA}
	\end{figure}

	The proposed SLB-NF-NOMA framework \textcolor[rgb]{0.00,0.00,0.00}{in Section II requires that} the users in the same cluster must be located in similar angular directions. However, it is not {always} applicable \textcolor[rgb]{0.00,0.00,0.00}{to practical} systems. \textcolor[rgb]{0.00,0.00,0.00}{This is because} the analog beamwidth is narrow in NFC systems, the users with different angular directions can not be covered by \textcolor[rgb]{0.00,0.00,0.00}{an analog beamformer merely} focusing on one specific location. Hence, in this section, \textcolor[rgb]{0.00,0.00,0.00}{we continue to} propose a MLB-NF-NOMA framework, as shown in Fig.~\ref{NF_MABC_NOMA}. MLB-NF-NOMA is able to generate multiple {sub-analog-beamformers} to serve the users with different angular directions simultaneously. \textcolor[rgb]{0.00,0.00,0.00}{To realize this multiple-location beamforcusing, we employ the beam-splitting technique~\cite{Wei2018MultiBeamNF}, where the adjacent antennas are separated to form two sub-arrays and each sub-array creates an sub-analog-beamformer, i.e., one sub-analog-beamformer focuses on the H-QoS user and the other sub-analog-beamformer focuses on the L-QoS user. Let $N_{m,h}$ and $N_{m,l}$ denote the number of antennas allocated to user $\mathcal{U} \left( m,h \right) 
		$ and user $\mathcal{U} \left( m,l \right) 
		$, respectively, where $N_{m,h}+N_{m,l}=N_{\mathrm{T}}$. Let $\mathbf{w}_{m,h}^{\mathrm{A}}\left( N_{m,h},r_{m,h},\theta _{m,h} \right) $ and $\mathbf{w}_{m,l}^{\mathrm{A}}\left( N_{m,l},r_{m,l},\theta _{m,l} \right) $ denote the sub-analog-beamformers for user $\mathcal{U} \left( m,h \right) $ with $N_{m,h}$ antenna subarray and user $\mathcal{U} \left( m,l \right) $ with $N_{m,l}$ antenna subarray, respectively. As a result, the analog beamformer for cluster $m$ can be formulated as follows
		\begin{equation}\label{multi_beam}
			\setlength{\abovedisplayskip}{5pt}
			\setlength{\belowdisplayskip}{5pt}
			\mathbf{w}_{m}^{\mathrm{A}}=\left[ \begin{array}{c}
				\mathbf{w}_{m,h}^{\mathrm{A}}\left( N_{m,h},r_{m,h},\theta _{m,h} \right)\\
				\mathbf{w}_{m,l}^{\mathrm{A}}\left( N_{m,l},r_{m,l},\theta _{m,l} \right)\\
			\end{array} \right] .
	\end{equation}}
	
	The achievable rate of H-QoS user $\mathcal{U} \left( m,h \right) 
	$ and L-QoS user $\mathcal{U} \left( m,l \right) 
	$ in MLB-NF-NOMA can be obtained by replacing the analog beamformer $\mathbf{w}_{m}^{\mathrm{A}}$ in~\eqref{gamma_11} and~\eqref{RLL} by~\eqref{multi_beam}.
	\subsection{Problem Formulation }
	Given the proposed MLB-NF-NOMA framework, we \textcolor[rgb]{0.00,0.00,0.00}{still} aim for maximizing the \textcolor[rgb]{0.00,0.00,0.00}{sum rate of H-QoS users} based on designing antenna allocation, hybrid beamforming \textcolor[rgb]{0.00,0.00,0.00}{design} and power allocation, subject to QoS requirements of \textcolor[rgb]{0.00,0.00,0.00}{all users}. The \textcolor[rgb]{0.00,0.00,0.00}{corresponding} optimization problem can be \textcolor[rgb]{0.00,0.00,0.00}{formulated} as follows: 
	\begin{subequations}\label{OP_multiBF}
		\setlength{\abovedisplayskip}{5pt}
		\setlength{\belowdisplayskip}{5pt}
		\begin{align}
			&\underset{\mathbf{w}_{m}^{\mathrm{A}},\mathbf{w}_{m}^{\mathrm{D}},p_{m,k}>0,N_{m,k}>0}{\max}\sum_{m\in \mathcal{M}}{R_{m,h}}	\\
			&\mathrm{s}.\mathrm{t}.~R_{m,k}\geqslant R_{m,k}^{\min},\label{OP_multiBF:b} \\ 
			&   \ \ \ \ \	\sum_{m=1}^{M_{\mathrm{RF}}}{\sum_{k\in \left\{ h,l \right\}}{p_{m,k}}}\leqslant P_{\max},\label{OP_multiBF:c} \\ 	
			&   \ \ \ \ \ \left| \left[ \mathbf{w}_{m}^{\mathrm{A}} \right] _n \right|=\frac{1}{\sqrt{N_{\mathrm{T}}}},\label{OP_multiBF:d} \\ 
			&   \ \ \ \ \ \left\| \mathbf{w}_{m}^{\mathrm{D}} \right\| _2=1	,\label{OP_multiBF:e} \\ 	
			&   \ \ \ \ \ N_{m,h}+N_{m,l}=N_{\mathrm{T}}, \label{OP_multiBF:f} \\
			&   \ \ \ \ \ N_{m,h},N_{m,l}\geqslant N_{\min}, \label{OP_multiBF:g} \\
			&   \ \ \ \ \ N_{m,h},N_{m,l}\in \mathbb{Z} ^+ 
			, \label{OP_multiBF:h}       
		\end{align} 
	\end{subequations}
	where $R_{m,h}$ and $R_{m,l}$ are defined as in~\eqref{gamma_11} and~\eqref{gamma_22}, $N_{\min}$ is the minimum number of antennas allocated to each user $k\in \left\{ h,l \right\} $, and $m\in \mathcal{M}$.
	
	{Compared to the formulated optimization problem~\eqref{OP_singleBF} for SLB-NF-NOMA, the main challenges of the considered problem~\eqref{OP_multiBF} are summarized in the following two aspects. First, due to the integer contains in~\eqref{OP_multiBF:f}-\eqref{OP_multiBF:h}, problem~\eqref{OP_multiBF} is a mixed-integer nonlinear programming problem, which is generally difficult to solve. The exhaustive search method can be applied to find the optimal $\left\{ N_{m,h},N_{m,l} \right\} 
		$. However, the computational complexity is prohibitively high. Second, the analog and digital beamformer, and the transmit power are highly coupled, the objective function and constraint~\eqref{OP_multiBF:b} are non-convex. Therefore, the considered problem~\eqref{OP_multiBF} for MLB-NF-NOMA is more challenging to solve than that for SLB-NAF-NOMA.}
	\vspace{-0.5cm}
	\subsection{Proposed Solution}
	{In this section, we propose a four-step algorithm for solving the optimization problem~\eqref{OP_multiBF}. Firstly, a novel multi-location-focused analog beamformer design method is proposed. Then, SVD-ZF digital beamformer is developed for reducing the inter-cluster interference. After that, a new antenna allocation algorithm is proposed based on many-to-one matching. Finally, a suboptimal power allocation algorithm is proposed based on fractional programming. }
	\subsubsection{\textcolor[rgb]{0.00,0.00,0.00}{Multiple-Location-Focused Analog Beamformer Design}} 
	In MLB-NF-NOMA, to enable analog beamfomer focusing energy on users with arbitrary locations, the sub-analog-beamformers $\mathbf{w}_{m,h}^{\mathrm{A}}\left( N_{m,h},r_{m,h},\theta _{m,h} \right) $ and $\mathbf{w}_{m,l}^{\mathrm{A}}\left( N_{m,l},r_{m,l},\theta _{m,l} \right)$ for H-QoS user $\mathcal{U} \left( m,h \right) 
	$ and L-QoS user $\mathcal{U} \left( m,l \right) 
	$ can be designed as follows
	\begin{equation}\label{analog_beam_m1}
		\mathbf{w}_{m,h}^{\mathrm{A}}\left( N_{m,h},r_{m,h},\theta _{m,h} \right) 
		=\frac{1}{\sqrt{N_{\mathrm{T}}}}\left[ e^{-j\frac{2\pi}{\lambda}r_{m,h}^{\left( 0 \right)}},e^{-j\frac{2\pi}{\lambda}r_{m,h}^{\left( 1 \right)}}\cdots ,e^{-j\frac{2\pi}{\lambda}r_{m,h}^{\left( N_{m,h}-1 \right)}} \right] ^T.
	\end{equation}
	\begin{equation}\label{analog_beam_m2}
		\mathbf{w}_{m,l}^{\mathrm{A}}\left( N_{m,l},r_{m,l},\theta _{m,l} \right) 
		=\frac{1}{\sqrt{N_{\mathrm{T}}}}\left[ e^{-j\frac{2\pi}{\lambda}r_{m,l}^{\left( N_{m,h} \right)}},e^{-j\frac{2\pi}{\lambda}r_{m,l}^{\left( N_{m,h}+1 \right)}}\cdots ,e^{-j\frac{2\pi}{\lambda}r_{m,l}^{\left( N_{\mathrm{T}}-1 \right)}} \right] ^T .
	\end{equation}
	
	Finally, we obtain the \textcolor[rgb]{0.00,0.00,0.00}{multiple-location-focused analog beamformer $\mathbf{w}_{m}^{\mathrm{A}}$} for cluster $m$ via~\eqref{multi_beam}.
	\begin{remark}
		Different from the \textcolor[rgb]{0.00,0.00,0.00}{single-location-focused analog beamformer} in~\eqref{single_beamfocusing}, which only depends on the H-QoS user's location, the \textcolor[rgb]{0.00,0.00,0.00}{multiple-location-focused analog beamformer} obtained via~\eqref{multi_beam} depends not only on the H-QoS user's location but also on the L-QoS user's location. In addition, the dimension of the two {sub-analog-beamformers} in~\eqref{analog_beam_m1} and~\eqref{analog_beam_m2} are determined by the number of antennas allocated to the H-QoS user and L-QoS user, which provides a new DoF for {MLB-NF-NOMA} design.
	\end{remark}
	
\begin{figure}[t]
		\setlength{\belowcaptionskip}{-20pt}
		\centering
		\includegraphics[scale=0.6]{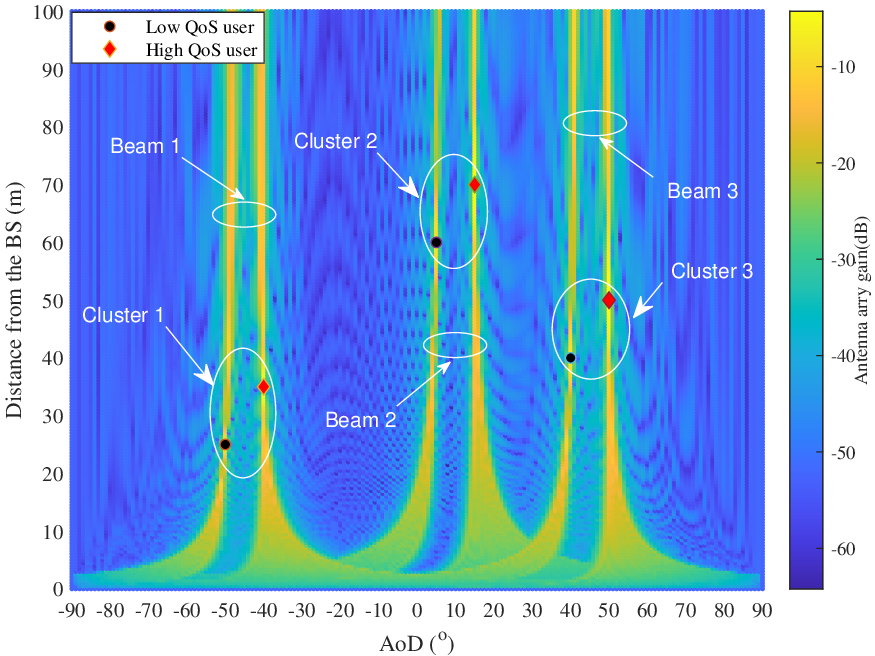}
		\caption{\justify{\normalsize Antenna array gain map for MLB-NF-NOMA system. We assume $N_{\mathrm{T}}=1024$, the locations $\left( r_{m,h},\theta _{m,h} \right)$ and $\left( r_{m,l},\theta _{m,l} \right)$ of H-QoS and L-QoS users in cluster 1, cluster 2, and cluster 3 are $\left( 25\mathrm{m},-50^{\mathrm{o}} \right) $ and $\left( 35\mathrm{m},-40^{\mathrm{o}} \right) $, $\left( 60\mathrm{m},5^{\mathrm{o}} \right) $ and $\left( 70\mathrm{m},15^{\mathrm{o}} \right) $, $\left( 40\mathrm{m},40^{\mathrm{o}} \right)$ and $\left( 50\mathrm{m},50^{\mathrm{o}} \right)$, respectively.} }
		\label{multi_Beam}
	\end{figure}

	For MLB-NF-NOMA, the analog beamformer~\eqref{multi_beam} generated by the beam-splitting design will be focused on the location $\left( r_{m,h},\theta _{m,h} \right) $ and $\left( r_{m,l},\theta _{m,l} \right) $ simultaneously. To illustrate the effect of the \textcolor[rgb]{0.00,0.00,0.00}{multiple-location-focused analog beamformer}, we plot the antenna array gain map for {MLB-NF-NOMA} in \textcolor[rgb]{0.00,0.00,0.00}{Fig.~\ref{multi_Beam}}. As observed in \textcolor[rgb]{0.00,0.00,0.00}{Fig.~\ref{multi_Beam}}, in each cluster, the generated two {sub-analog-beamformers} can simultaneously focus most beam energy around the desired H-QoS and L-QoS users' locations, while such \textcolor[rgb]{0.00,0.00,0.00}{multiple-location} energy focusing ability \textcolor[rgb]{0.00,0.00,0.00}{cannot be realized} by \textcolor[rgb]{0.00,0.00,0.00}{single-location-focused analog beamformer design} in {SLB-NF-NOMA} and traditional multi-beam scheme in FF-NOMA~\cite{Wei2018MultiBeamNF}.  
	
	\textcolor[rgb]{0.00,0.00,0.00}{As} the \textcolor[rgb]{0.00,0.00,0.00}{multiple-location-focused analog beamformer} \textcolor[rgb]{0.00,0.00,0.00}{is} formed via \textcolor[rgb]{0.00,0.00,0.00}{the} beam-splitting technology, there are some beamfocusing gain losses compared to \textcolor[rgb]{0.00,0.00,0.00}{the single-location-focused analog beamformer}, which is \textcolor[rgb]{0.00,0.00,0.00}{explained} in the following \textbf{Lemma~\ref{inequialty_Gain}}.
	\begin{lemma}\label{inequialty_Gain}
		Let $\mathbf{w}_{m}^{\mathrm{SA}}$ and $\mathbf{w}_{m}^{\mathrm{MA}}
		$ denote \textcolor[rgb]{0.00,0.00,0.00}{single- and multiple-location-focused analog beamformer obtained via}~\eqref{single_beamfocusing} and~\eqref{multi_beam}, respectively. we have the following inequality: 
		\begin{equation}\label{relationship}
			\setlength{\abovedisplayskip}{5pt}
			\setlength{\belowdisplayskip}{5pt}	
			\underset{k\in \left\{ h,l \right\}}{\max}\left| \mathbf{b}^H\left( r_{m,k},\theta _{m,k} \right) \mathbf{w}_{m}^{\mathrm{MA}} \right|\leqslant \left| \mathbf{b}^H\left( r_{m,h},\theta _{m,h} \right) \mathbf{w}_{m}^{\mathrm{SA}} \right|.
		\end{equation}	
		Proof: Please refer to Appendix~A.
		
	\end{lemma}
	
	Thus, compared to the \textcolor[rgb]{0.00,0.00,0.00}{single--location-focused analog beamformer}, \textcolor[rgb]{0.00,0.00,0.00}{it can be found that} some beamfocusing gains are sacrificed by \textcolor[rgb]{0.00,0.00,0.00}{multiple-location-focused analog beamformer}. \textcolor[rgb]{0.00,0.00,0.00}{Nevertheless}, \textcolor[rgb]{0.00,0.00,0.00}{multiple-location-focused analog beamformer design} \textcolor[rgb]{0.00,0.00,0.00}{can perform NOMA transmission for the users with arbitrary locations, which provides a higher flexibility for applying NOMA in NFC. }
	\subsubsection{SVD-ZF Digital Beamformer Design}
	In the considered {MLB-NF-NOMA} system, the H-QoS user and the L-QoS user have different angular directions. As a result, the \textcolor[rgb]{0.00,0.00,0.00}{beamspace} channel correlation between the two users may be not high enough. \textcolor[rgb]{0.00,0.00,0.00}{In} this case, SVD-ZF digital beamformer design~\cite{Wang2017SpectrumAE} is \textcolor[rgb]{0.00,0.00,0.00}{employed} in this paper. Specifically, define the \textcolor[rgb]{0.00,0.00,0.00}{beamspace channel} matrix of all users in the $m$-th cluster as $\widetilde{\mathbf{G}}_m=\left[ \widetilde{\mathbf{g}}_{m,h},\widetilde{\mathbf{g}}_{m,l} \right] 
	$. First, performing SVD on $\widetilde{\mathbf{G}}_{m}^{T}
	$, we have
	\begin{equation}\label{SVD_decompose}
		\widetilde{\mathbf{G}}_{m}^{T}=\mathbf{U}_m\mathbf{\Lambda }_m\mathbf{V}_{m}^{H},
	\end{equation}
	where $\mathbf{U}_m$ is the left singular matrix, $\mathbf{\Lambda }_m$ is the singular value matrix with its diagonal entries sorted in descending order, and $\mathbf{V}_{m}$ is the right singular matrix.
	
	Then, we can obtain the equivalent channel vector of the $m$-th cluster via $\overline{\mathbf{g}}_m=\widetilde{\mathbf{G}}_m\mathbf{u}_{m,1}$, where $\mathbf{u}_{m,1}$ is the first column of $\mathbf{U}_{m}$. After that, the digital beamformer matrix $\overline{\mathbf{W}}^{\mathrm{D}}$ can be expressed as
	\begin{equation}\label{SVD_DBM}
		\overline{\mathbf{W}}^{\mathrm{D}}=\left[ \overline{\mathbf{w}}_{1}^{\mathrm{D}},\overline{\mathbf{w}}_{2}^{\mathrm{D}},\cdots ,\overline{\mathbf{w}}_{M_{\mathrm{RF}}}^{\mathrm{D}} \right] =\overline{\mathbf{G}}\left( \overline{\mathbf{G}}^H\overline{\mathbf{G}} \right) ^{-1},
	\end{equation} 
	where $\overline{\mathbf{G}}=\left[ \overline{\mathbf{g}}_1,\overline{\mathbf{g}}_2,\cdots ,\overline{\mathbf{g}}_{M_{\mathrm{RF}}} \right] $ is the equivalent channel matrix. 
	
	Similarly \textcolor[rgb]{0.00,0.00,0.00}{to}~\eqref{ZF_DB}, by normalizing the vectors $\left\{ \overline{\mathbf{w}}_{m}^{\mathrm{D}} \right\} 
	$, we can obtain the digital beamformer of the $m$-th cluster as follows
	\begin{equation}\label{SVD_DB}
		\mathbf{w}_{m}^{\mathrm{D}}=\frac{\overline{\mathbf{w}}_{m}^{\mathrm{D}}}{\left\| \overline{\mathbf{w}}_{m}^{\mathrm{D}} \right\| _2}.
	\end{equation} 
	\subsubsection{Many-to-One Matching Based Antenna Allocation}
	{For given analog and digital beamformer, and power allocation, the antenna allocation problem can be formulated as follows}
	\begin{subequations}\label{Antenna_Allocation}
		\setlength{\abovedisplayskip}{5pt}
		\setlength{\belowdisplayskip}{5pt}
		\begin{align}
			&\underset{N_{m,k}>0}{\max}~\sum_{m\in \mathcal{M}}{R_{m,h}} \\
			&\mathrm{s}.\mathrm{t}.~\eqref{OP_multiBF:d}-\eqref{OP_multiBF:f}.    
		\end{align} 
	\end{subequations}
	
	Before solving problem~\eqref{Antenna_Allocation}, we define the set of all possible antenna allocation strategy as
	\begin{equation}\label{all_possible_AA}
		\mathcal{Q} =\left\{ \left( N_{m,h},N_{m,l} \right) |\begin{array}{l}
			N_{m,h}=N_{\mathrm{T}}-\left( N_{\min}+q \right) ,\\
			N_{m,l}=N_{\min}+q,\\
			q=0,1,\cdots ,Q\\
		\end{array} \right\}, 
	\end{equation} 
	where $\mathcal{Q} _q\in \mathcal{Q} $ is the $q$-th antenna allocation strategy, $\left| \mathcal{Q} \right|=Q$ and $Q=N_{\mathrm{T}}-2N_{\min}$ is the total number of possible antenna allocation strategies.
	
	The above problem can be solved by many-to-one matching with two sides, i.e., antenna allocation strategies and clusters. Combining with the antenna allocation problem, define the many-to-one matching function $\varPsi $ as~\cite{Cui2018UnsupervisedML,Zhao2017SpectrumAA} 
	\begin{enumerate}[(1)]
		\item $\left| {\varPsi  \left( m \right)} \right| = 1$, for each cluster $\forall m \in {\cal M}$, $\varPsi  \left( m \right) \in {\cal Q}$,
		%\item $\left| \varPsi \left( \mathcal{Q} _q \right) \right|\leqslant M_q$ for each antenna allocation strategy $\mathcal{Q} _q$;	
		\item $\varPsi  \left( m \right){\rm{ = }}\mathcal{Q} _q$ if and only if $m \in \varPsi  \left( \mathcal{Q} _q \right)$,
	\end{enumerate}
	where definition \textcolor[rgb]{0.00,0.00,0.00}{(1)} means that each cluster can only be matched with one antenna allocation strategy. Definition \textcolor[rgb]{0.00,0.00,0.00}{(2)} implies that if cluster $m$ is matched with antenna allocation strategy $\mathcal{Q} _q$, then antenna allocation strategy $\mathcal{Q} _q$ is also matched with cluster $m$.
	
	We assume that each player have preferences over the players of the other set. The preference of cluster $m$ and antenna allocation strategy $\mathcal{Q} _q$ is based on the preference function, which is defined as follows 
	\begin{equation}\label{preference}
		\varPhi _{q,m}=\left. \frac{\sum_{k\in \left\{ h,l \right\}}{p_{m,k}\left| \mathbf{g}_{m,k}^{H}\mathbf{w}_{m}^{\mathrm{A}} \right|^2}}{P_m\sum_{i\ne m,i\in \mathcal{M}}{\sum_{k\in \left\{ h,l \right\}}{\left| \mathbf{g}_{i,k}^{H}\mathbf{w}_{m}^{\mathrm{A}} \right|^2}}} \right|_{\left( N_{m,h},N_{m,l} \right) =\mathcal{Q} _q}.
	\end{equation} 
	%where the analog beamformer $\mathbf{w}_{m}^{\mathrm{A}}$ can be obtained in the following
	
	Define the utility of cluster $m$ utilizing antenna allocation strategy $\mathcal{Q} _q$ as {follows }
	\begin{equation}\label{Uqm}
		\setlength{\abovedisplayskip}{5pt}
		\setlength{\belowdisplayskip}{5pt}
		U_m=R_{m,h}-\frac{P_m}{\eta _m}\sum_{i\ne m,i\in \mathcal{M}}{\sum_{k\in \left\{ h,l \right\}}{\left| g_{m,\left( i,k \right)} \right|^2}},
	\end{equation} 
	where $\eta _m$ is a scale coefficient.
	
	The utility function of the $q$-th antenna allocation strategy is defined as {follows}
	\begin{equation}\label{Uq}
		\setlength{\abovedisplayskip}{5pt}
		\setlength{\belowdisplayskip}{5pt}
		U_q=\sum_{m\in \varPsi \left( \mathcal{Q} _q \right)}{R_{m,h}}+\sum_{m\in \varPsi \left( \mathcal{Q} _{-q} \right)}{R_{m,h}},
	\end{equation}
	where $\varPsi \left( \mathcal{Q} _{-q} \right) $ denote the set of clusters allocated with other antenna allocation strategies.
	
	Since the utility of cluster $m$ depends not only on its own channel but also on the the other clusters' antenna allocation strategies. To tackle this interdependence, we utilize swap operations between any two clusters to exchange their allocated antenna allocation strategies. First, define swap matching~\cite{Zhao2017SpectrumAA,Cui2018UnsupervisedML,9167258} as follows
	\begin{equation}\label{define swap matching}
		\setlength{\abovedisplayskip}{3pt}
		\setlength{\belowdisplayskip}{3pt}
		\varPsi _{m}^{\widetilde{m}}=\left\{ \varPsi \backslash \left\{ \left( m,\mathcal{Q} _q \right) ,\left( \widetilde{m},\mathcal{Q} _{\widetilde{q}} \right) \right\} \cup \left\{ \left( \widetilde{m},\mathcal{Q} _q \right) ,\left( m,\mathcal{Q} _{\widetilde{q}} \right) \right\} \right\},
	\end{equation}
	where $\varPsi \left( m \right) =\mathcal{Q} _q,\varPsi \left( \widetilde{m} \right) =\mathcal{Q} _{\widetilde{q}}$.
	
	The swap matching enables cluster \emph{m} and cluster ${\widetilde m}$ to switch their assigned antenna allocation strategies. Then, we introduce the definition of swap-blocking pair. Given a matching function $\varPsi$ and assume that $\varPsi \left( m \right) =\mathcal{Q} _q$ and $\varPsi \left( \widetilde{m} \right) =\mathcal{Q} _{\widetilde{q}}$, a pair of clusters $\left( {m,\widetilde m} \right)$ is a swap-blocking pair if and only if
	\begin{enumerate}[(1)]
		\item $\forall x\in \left\{ m,\widetilde{m},q,\widetilde{q} \right\} $, $U_x\left( \varPsi _{m}^{\widetilde{m}} \right) \ge U_{\omega}\left( x \right) 
		$;
		\item $\exists x\in \left\{ m,\widetilde{m},q,\widetilde{q} \right\} 
		$, $U_x\left( \varPsi _{m}^{\widetilde{m}} \right) >U_x\left( \varPsi \right) 
		$,
	\end{enumerate}
	where $U_x\left( \varPsi \right) $ and $U_x\left( \varPsi _{m}^{\widetilde{m}} \right)$ are the utilities of player $x$ (cluster $x$ or the $x$-th antenna allocation strategy), under the matching state $\varPsi$ and $U_x\left( \varPsi _{m}^{\widetilde{m}} \right)$, respectively.
	
	According to the above definition, it is noted that if two clusters want to switch their assigned antenna allocation strategies, both of the conditions should be satisfied. Condition \textcolor[rgb]{0.00,0.00,0.00}{(1)} indicates that all the involved players' utilities should not be reduced after the swap operation. Condition \textcolor[rgb]{0.00,0.00,0.00}{(2)} indicates that after the swap operation, at least one of the players' utilities is increased.
	
	Based on the above analysis, the proposed antenna assignment algorithm is summarized in \textbf{Algorithm~\ref{AAA}}. There are two processes in \textbf{Algorithm~\ref{AAA}} as follows
	\begin{enumerate}[(1)]
		\item \textbf{Initialization Process}: \textcolor[rgb]{0.00,0.00,0.00}{Let $\mathcal{Q} _q$ and ${{\cal M}_{\rm un}}$ denote the set of clusters assigned to the antenna allocation strategy and the set of clusters that are not matched with any antenna allocation strategy, respectively}. \textcolor[rgb]{0.00,0.00,0.00}{Let ${\cal M}_q^{\rm PRO}$ denote the set of clusters that propose to antenna allocation strategy $\mathcal{Q} _q$.} During the matching period, each un-matched cluster proposes to the antenna allocation strategy that can provide the highest preference function value and has never rejected it before. Then, each antenna allocation strategy accepts the proposal with the highest preference function value it can provide and rejects other clusters. Repeat the above process until the set of un-matched clusters is empty.
		\item \textbf{Swapping Process}: Swap operations among clusters are enabled to further improve the performance of the antenna allocation algorithm. With the obtained cluster set ${\cal M}_q$, $q\in \left\{ 1,\cdots, Q \right\}$, in the \textbf{Initialization Process}, each cluster tries to search for another cluster to construct the swap-blocking pair and update their corresponding matching state and cluster set ${\cal M}_q$, $q\in \left\{ 1,\cdots, Q \right\}$. This operation will continue until there is no swap-blocking pair.
	\end{enumerate}

	\begin{algorithm}[!t]
		\caption{Antenna allocation Algorithm}
		\label{AAA}
		\begin{algorithmic}[1]
			\STATE \textbf{Initialization Process}:
			\STATE   Initialize ${{\cal M}_q} = \emptyset$, ${\cal
				M}_q^{\rm PRO} = \emptyset$, and ${{\cal M}_{\rm un}} = {\cal M}$, $q\in \left\{ 1,2,\cdots Q \right\} 
			$.
			\WHILE {${{\cal M}_{\rm NOT}} \ne \emptyset$}
			\STATE the un-matched cluster $m\in \left\{ \mathcal{M} _{\mathrm{un}}\setminus \cup _{q\in \left\{ 1,\cdots ,Q \right\}}\mathcal{M} _q \right\} 
			$ proposes to choose its best antenna allocation strategy $\mathcal{Q} _{q^*}$, where $q^*=\underset{q\in \left\{ 1,\cdots, Q \right\}}{\max}\varPhi _{q,m}
			$;
			\STATE  update ${\cal M}_q^{\rm PRO}$ based on the results obtained from the last step, i.e., $ \mathcal{M} _{q}^{\mathrm{PRO}}=\mathcal{M} _{q}^{\mathrm{PRO}}\cup \left\{ m \right\} 
			$;
			\STATE  update set $\mathcal{M} _q=\mathcal{M} _q\cup \mathcal{M} _{q}^{\mathrm{PRO}}$, and antenna allocation strategy $\mathcal{Q} _{q}$ accepts all the clusters in ${{\cal M}_q}$;
			\ENDWHILE
			\STATE \textbf{Swapping Process}:
			\STATE \textbf{ repeat}
			\STATE   For any cluster $m \in {{\cal M}_q}$, it searches for another cluster $\widetilde m \in {{\cal M}_{\widetilde q}}$, where
			$\widetilde q \ne q, q\in \left\{ 1,\cdots, Q \right\} $.
			\IF {cluster pair $\left( {m, \widetilde m} \right)$ is a swap-blocking pair}
			\STATE   update $\varPsi \left( m \right) =\mathcal{Q} _{\widetilde{q}}$ and $\varPsi \left( \widetilde{m} \right) =\mathcal{Q} _q$; 
			\ELSE
			\STATE   keep the current matching state unchanged;
			\ENDIF
			\STATE  \textbf{until} No swap-blocking pair can be found.
			\STATE  \textbf{Output}: all the matched cluster ${{\cal M}_q}, q\in \left\{ 1,\cdots, Q \right\}$.
		\end{algorithmic}
	\end{algorithm}
	\subsubsection{Power Allocation}
	Given the antenna allocation, analog beamformer and digital beamformer, the power allocation problem can be formulated as
	\begin{subequations}\label{OP_multiBF_power_allocation}
		\setlength{\abovedisplayskip}{5pt}
		\setlength{\belowdisplayskip}{5pt}
		\begin{align}
			&\underset{p_{m,h},p_{m,l}>0}{\max}~\sum_{m\in \mathcal{M}}{R_{m,h}} \\
			&\mathrm{s}.\mathrm{t}.~p_{m,h}\geqslant \frac{r_{m,h}^{\min}\left( I_{m,h}^{\mathrm{inter}}\left( \mathbf{p}_{-m} \right) +\sigma ^2 \right)}{\left| g_{m,h} \right|^2},\label{OP_multiBF_power_allocation:b} \\ 
			&   \ \ \ \ \ p_{m,l}\geqslant r_{m,l}^{\min}\left( p_{m,h}+\frac{I_{m,l}^{\mathrm{inter}}\left( \mathbf{p}_{-m} \right) +\sigma ^2}{\left| g_{m,l} \right|^2} \right) ,\label{OP_multiBF_power_allocation:c} \\ 	
			&   \ \ \ \ \ p_{m,l}\geqslant r_{m,l}^{\min}\left( p_{m,h}+\frac{I_{m,h}^{\mathrm{inter}}\left( \mathbf{p}_{-m} \right) +\sigma ^2}{\left| g_{m,h} \right|^2} \right) ,\label{OP_multiBF_power_allocation:d} \\  
			&   \ \ \ \ \  \eqref{OP_multiBF:c}.\label{power constarint}	      
		\end{align} 
	\end{subequations}
	
	However, problem~\eqref{OP_multiBF_power_allocation} is difficult to solve due to the nonconvex form of the weighted sum rate objective function. According to fractional programming~\cite{Shen2018FractionalPF}, we introduce the auxiliary variables $\left\{ \beta _m \right\} $ and applying quadratic transform to each signal-to-interference-plus-noise-ratio (SINR) term, the objective function can be transformed into solvable formula. Particular, the objective function in~\eqref{OP_multiBF_power_allocation} is reformulated as
	\begin{equation}\label{obj_rewritten}
		R_{\mathrm{sum}}\left( \mathbf{p},\boldsymbol{\beta } \right) =\sum_{m\in \mathcal{M}}{\log _2\left( 1+2\beta _m\sqrt{p_{m,h}\left| g_{m,h} \right|^2}-\beta _{m}^{2}\left( I_{m,h}^{\mathrm{inter}}\left( \mathbf{p}_{-m} \right) +\sigma ^2 \right) \right)},
	\end{equation} 
	where $\boldsymbol{\beta }$ refers to the collection $\left\{ \beta _m \right\} $.
	
	For fixed power allocation strategy $\mathbf{p}
	$, the optimal $\beta _m$ is given by
	\begin{equation}\label{optimal_beta}
		\beta _m=\frac{\sqrt{p_{m,h}\left| g_{m,h} \right|^2}}{I_{m,h}^{\mathrm{inter}}\left( \mathbf{p}_{-m} \right) +\sigma ^2}.
	\end{equation} 
	
	Then, solving problem~\eqref{OP_multiBF_power_allocation} is equivalent to solving the following convex optimization problem  
	\begin{subequations}\label{OP_multiBF_power_allocation1}
		\setlength{\abovedisplayskip}{-9pt}
		\setlength{\belowdisplayskip}{3pt}
		\begin{align}
			&\underset{\mathbf{p},\boldsymbol{\beta }}{\max}~R_{\mathrm{sum}}\left( \mathbf{p},\boldsymbol{\beta } \right) \\
			&\mathrm{s}.\mathrm{t}.~\eqref{OP_multiBF_power_allocation:b}-\eqref{power constarint}.
		\end{align} 
	\end{subequations}
	
	The proposed iterative power allocation algorithm to solve problem~\eqref{OP_multiBF_power_allocation} is summarized 
	in \textbf{Algorithm~\ref{MBC_NOMA_NFC_power_algorithm}}.
	\begin{algorithm} 
		\caption{Proposed Power Allocation Algorithm}
		\label{MBC_NOMA_NFC_power_algorithm}
		\begin{algorithmic}[1]
			\STATE Initialize $\mathbf{p}$ to a feasible value		
			\REPEAT
			\STATE  update $\boldsymbol{\beta }$ via~\eqref{optimal_beta}; 
			\STATE   update $\mathbf{p}$ by solving problem~\eqref{OP_multiBF_power_allocation1} with for fixed $\boldsymbol{\beta }$; 
			\UNTIL {the objective value of problem~\eqref{OP_multiBF_power_allocation1} converges.}
		\end{algorithmic}
	\end{algorithm}  
	
	In \textbf{Algorithm~\ref{MBC_NOMA_NFC_power_algorithm}}, the initial feasible points $\left\{ p_{m,k} \right\} $ are needed, \textcolor[rgb]{0.00,0.00,0.00}{which can be difficult.} In the following, we formulate a feasible searching problem to find the feasible points $\left\{ p_{m,k} \right\} $. 
	\begin{subequations}\label{find_power}
		\setlength{\abovedisplayskip}{-5pt}
		\setlength{\belowdisplayskip}{5pt}
		\begin{align}
			& \mathrm{find}~\mathbf{p}  \\
			&\mathrm{s}.\mathrm{t}.\textcolor[rgb]{0.00,0.00,0.00}{~\eqref{OP_multiBF_power_allocation:b}-\eqref{power constarint}.}
		\end{align} 
	\end{subequations}
	
	Problem~\eqref{find_power} is convex, which can be can be solved efficiently by standard algorithms or \textcolor[rgb]{0.00,0.00,0.00}{CVX}\textbf{~\cite{cvx}}. 
	\subsubsection{Overall Algorithm and Complexity Analysis}
	Based on the above \textcolor[rgb]{0.00,0.00,0.00}{steps}, the complete algorithm to
	realize the antenna allocation, hybrid beamforming and power allocation for \textcolor[rgb]{0.00,0.00,0.00}{solving problem~\eqref{OP_multiBF}} in {MLB-NF-NOMA} is \textcolor[rgb]{0.00,0.00,0.00}{summarized} in \textbf{Algorithm~\ref{MBC_NOMA_NFC_algorithm}}.
	\begin{algorithm}
		\caption{Proposed algorithm for\textbf{ MLB-NF-NOMA} system}
		\label{MBC_NOMA_NFC_algorithm}
		\begin{algorithmic}[1]
			\STATE  \textbf{Antenna Allocation}
			\STATE    ~~The antenna allocation strategy for each cluster is obtained via \textbf{Algorithm~\ref{AAA}};
			\STATE  \textbf{Multiple-Location-Focused Analog Beamformer Design}
			\STATE    ~~The analog beamformer $\mathbf{w}_{m}^{\mathrm{A}}$ is calculated via~\eqref{multi_beam}, $m\in \mathcal{M}$;		
			\STATE  \textbf{Digital Beamformer Design}
			\STATE  ~~Calculate the digital beamformer $\mathbf{w}_{m}^{\mathrm{D}}$ according to~\eqref{SVD_DB}, $m\in \mathcal{M}$;
			\STATE  \textbf{Power Allocation}
			\STATE  ~~Obtain power allocation strategy $\mathbf{p}$ via \textbf{Algorithm~\ref{MBC_NOMA_NFC_power_algorithm}}.
		\end{algorithmic}
	\end{algorithm} 
	
	\textcolor[rgb]{0.00,0.00,0.00}{The main complexity of \textbf{Algorithm~\ref{MBC_NOMA_NFC_algorithm}} for solving problem~\eqref{OP_multiBF} comes from \textbf{Algorithm~\ref{AAA}} and \textbf{Algorithm~\ref{MBC_NOMA_NFC_power_algorithm}}.} The proposed many-to-one matching based antenna allocation algorithm converges to a two-sided stable matching within a limited number of iterations, which has been proved in~\cite{9167258,Zhao2017SpectrumAA}. The complexity of \textbf{Algorithm~\ref{AAA}} \textcolor[rgb]{0.00,0.00,0.00}{in step 2} mainly lies in the number of the user proposing and swap operations. For the worst case, the proposing number is $2MQ$ and the maximum number of swap operations is $4M^2Q^2$. In addition, \textbf{Algorithm~\ref{MBC_NOMA_NFC_power_algorithm}} \textcolor[rgb]{0.00,0.00,0.00}{in step 4} is guaranteed \textcolor[rgb]{0.00,0.00,0.00}{to converge} to a stationary point~\cite{Shen2018FractionalPF} of problem~\eqref{OP_multiBF_power_allocation}. 
	In \textbf{Algorithm~\ref{MBC_NOMA_NFC_power_algorithm}}, the computation complexity to solve problem~\eqref{OP_multiBF_power_allocation1} and~\eqref{find_power} are $\mathcal{O} \left( \left( 2M_{\mathrm{RF}} \right) ^{3.5} \right) 
	$ and $\mathcal{O} \left( \left( 2M_{\mathrm{RF}} \right) ^{3.5} \right)$, respectively. Thus the over all complexity of \textbf{Algorithm~\ref{MBC_NOMA_NFC_power_algorithm}} can be calculated as $\mathcal{O} \left( \left( 2M_{\mathrm{RF}} \right) ^{3.5}\left( t_{\max}+1 \right) \right)	$, where $t_{\max}$ is the number of iterations. \textcolor[rgb]{0.00,0.00,0.00}{Therefore, the total computational complexity of \textbf{Algorithm~\ref{MBC_NOMA_NFC_algorithm}} is $\mathcal{O} \left( 4M^2Q^2+\left( 2M_{\mathrm{RF}} \right) ^{3.5}\left( t_{\max}+1 \right) \right) 
		$.}
	\section{\textcolor[rgb]{0.00,0.00,0.00}{Numerical} Results}  
	In this section, we provide numerical results for characterizing the proposed NF-NOMA frameworks. In particularity, we assume that the BS employs a uniform linear array (ULA) positioned in the $xy$-plane, with the midpoint of the BS antenna located in $\left( 0,0 \right) $. To prevent a large beamforming gain loss caused by endfire beamforming, we only consider the users uniformly distributed in the $\frac{1}{3}$ cell, which means that the AoDs of the users range from $-60^{\text{o}}$ to $60^{\text{o}}$. The considered communication system is assumed to be operated at $30$ GHz carrier frequency {(i.e., $\lambda =1~\mathrm{cm}$)}. The spacing $d$ between adjacent BS antennas is set {to} $\frac{\lambda}{2}$. Without special statements, the number of the RF chains is set to $M_{\mathrm{RF}}=4$. All the $K$ $(K=M_{\mathrm{RF}})$ users are grouped into $4$ clusters, where there are one L-QoS \textcolor[rgb]{0.00,0.00,0.00}{user} and one H-QoS \textcolor[rgb]{0.00,0.00,0.00}{user} in each cluster. The noise power is set to $\sigma ^2=-90$~dBm. We set the minimum QoS requirement for user $\mathcal{U} \left( m,h \right) $ and user $\mathcal{U} \left( m,l \right) $
	as $R_{m,h}^{\min}=6$~bit/s/Hz. 
	\subsection{SLB-NF-NOMA} 
	The simulated geometry for {SLB-NF-NOMA } is set as follows: the AoDs of the users in the four clusters are randomly distributed in the angle ranges $\theta _{1,k}\in \left[ -40^{\mathrm{o}},-30^{\mathrm{o}} \right] $, $\theta _{2,k}\in \left[ -5^{\mathrm{o}},+5^{\mathrm{o}} \right] 
	$, $\theta _{4,k}\in \left[ 30^{\mathrm{o}},40^{\mathrm{o}} \right] $ and $\theta _{2,k}=\theta _{3,k} 
	$. In addition, the radii of the users in the corresponding clusters are randomly distributed in $r_{1,k}\in \left[ 30,50 \right] $ \text{m}, $r_{2,k}\in \left[ 35,55 \right] $ \text{m}, $r_{3,k}\in \left[60,80 \right]$ \text{m}, and $r_{4,k}\in \left[40,60 \right]$ \text{m}. Without loss of generality, we set the AoD difference of the two uses in each cluster to $0.1^{\mathrm{o}}$. For performance comparison, we consider the following two benchmark schemes:
	\begin{enumerate}[(1)]
		\item '\textbf{NF-OMA}' scheme, where the $2K$ users are randomly scheduled into two time slots. In each time slot, there are total $K$ users and each user can only be associated with at most one RF chain. %The single beamfocusing analog beamformer and ZF digital beamfomer design are adopted.  	
		\item '\textbf{FF-NOMA-OMA}' scheme, it is noted that the FF-NOMA system \textcolor[rgb]{0.00,0.00,0.00}{cannot distinguish} cluster $2$ and cluster $3$ in the same direction. To solve this problem, we schedule cluster $2$ and cluster $3$ into two time slots. Specifically, Cluster $1$, cluster $2$ and cluster $4$ perform FF-NOMA scheme in the first time slot, and Cluster $1$, cluster $3$ and cluster $4$ perform {FF-NOMA} scheme in the second time slot.  	
	\end{enumerate}	
	\subsubsection{Average Sum Rate Versus $P_{\max}$} 
\begin{figure}[t]
		\setlength{\belowcaptionskip}{-20pt}
		\centering
		\includegraphics[scale=0.6]{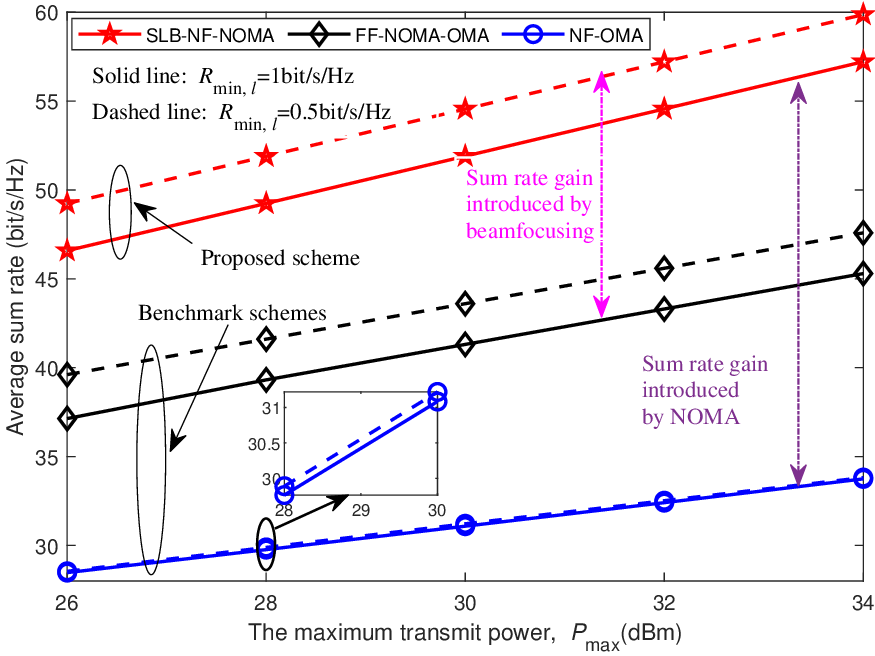}
		\caption{Average sum rate versus $P_{\max}$ for $N_{\mathrm{T}}=512$}% 
		\label{sumR_vs_Pmax}
	\end{figure}

	In Fig.~\ref{sumR_vs_Pmax}, we present the average sum rate achieved versus the maximum transmit power $P_{\max}$. It can be observed that the sum rate of all schemes increases upon increasing $P_{\max}$ and the proposed {SLB-NF-NOMA} scheme achieves the best performance. Moreover, the proposed {SLB-NF-NOMA} scheme outperforms the {FF-NOMA-OMA} scheme. This is because, NF-beamfocusing is able to distinguish the users with the similar angular direction and enhance the H-QoS users' effective channel gain even if they locate far from the BS, while such ability is not achievable for conventional FF-beamsteering. In addition, the sum rate enhancement attained by NOMA based schemes upon increasing $P_{\max}$ is more significant than OMA based scheme, since NOMA benefits from a flexible resource allocation scheme. It can
	also be observed that the sum rate achieved by the three schemes decreases as $R_{\min ,l}$ increases. This is indeed expected, since a higher QoS requires more transmit power to be allocated, thus degrading the sum rate. Additionally, in contrast to {SLB-NF-NOMA} and {FF-NOMA-OMA} schemes, the degradation of sum rate in NF-OMA scheme upon increasing $R_{\min ,l}$ becomes negligible. This is because NF-OMA scheme is interference-free, when $R_{\min ,l}$ becomes stricter, the rate requirement can be easily satisfied by increasing fewer transmit power.  
	\subsubsection{{Average Sum Rate} Versus $N_{\mathrm{T}}$}  
\begin{figure}[t]
		\setlength{\belowcaptionskip}{-20pt}
		\centering
		\includegraphics[scale=0.6]{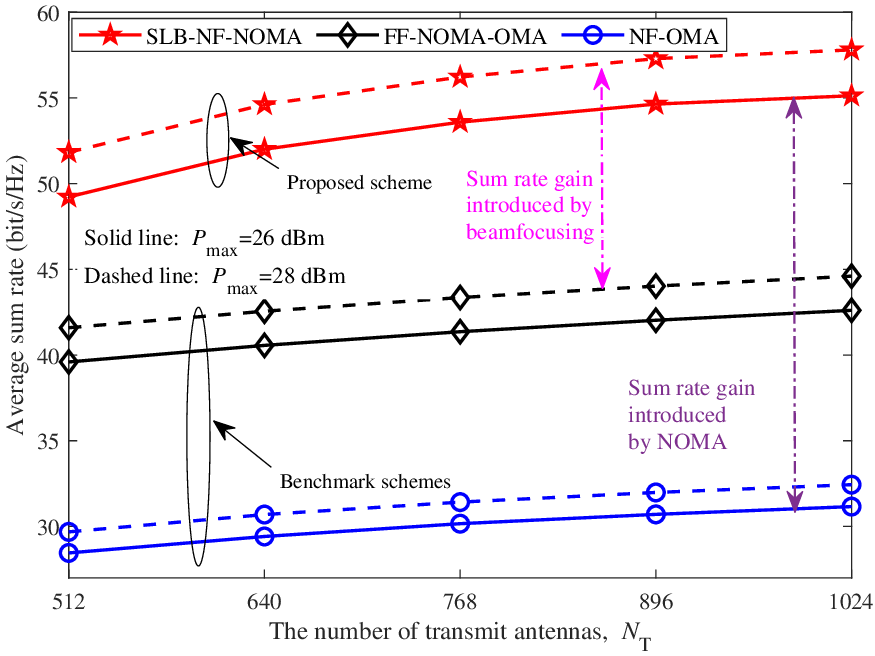}
		\caption{Average sum rate versus $N_{\mathrm{T}}$ for $R_{\min,l}=0.5$~bit/s/Hz} %
		\label{sumR_vs_NT}
	\end{figure}

	Fig.~\ref{sumR_vs_NT} illustrates the average sum rate versus the number of antennas $N_{\mathrm{T}}$. We observe that the average sum rate increases monotonically with $N_{\mathrm{T}}$ due to the increased array gain. Compared to the two benchmark schemes, a higher sum rate can be attained by the proposed {SLB-NF-NOMA} scheme due to its higher flexibility of exploiting NF-beamfocusing and NOMA.
	\subsubsection{Total Interference Versus $N_{\mathrm{T}}$} 
	\begin{figure}[t]
		\setlength{\belowcaptionskip}{-20pt}
		\centering
		\includegraphics[scale=0.6]{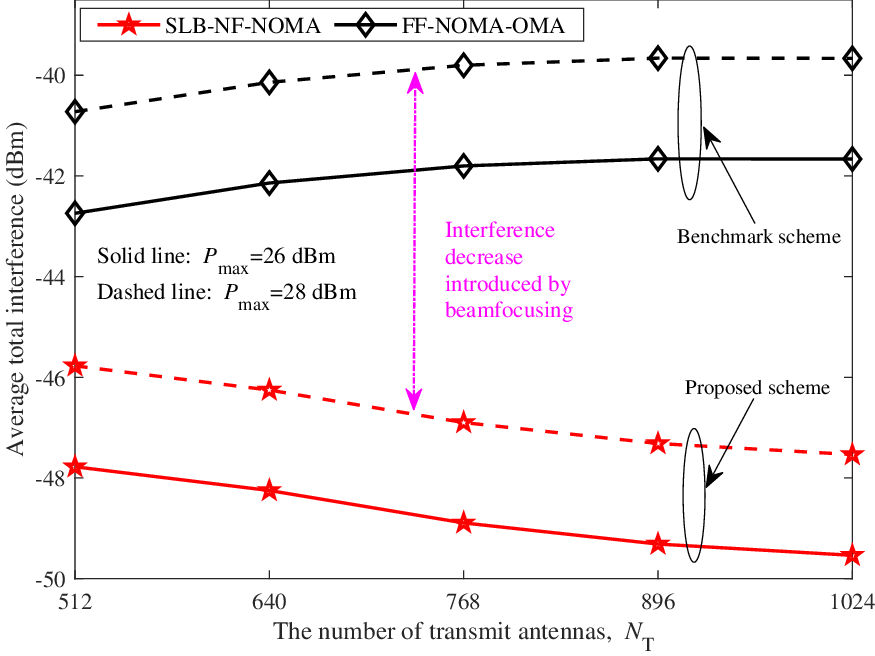}
		\caption{Average total interference versus $N_{\mathrm{T}}$ for $R_{\min,l}=0.5$~bit/s/Hz}%, for $P_{\max}=30$~dBm}
		\label{Interference_vs_NT}
	\end{figure}
	
	In Fig.~\ref{Interference_vs_NT}, we plot the average total interference versus $N_{\mathrm{T}}$. As shown in Fig.~\ref{Interference_vs_NT}, by increasing $N_{\mathrm{T}}$, the total interference decreases in the proposed {SLB-NF-NOMA scheme}, \textcolor[rgb]{0.00,0.00,0.00}{but} increases in the {FF-NOMA-OMA} scheme. The reason behind this can be explained as \textcolor[rgb]{0.00,0.00,0.00}{follows}. Since conventional FF-beamsteering \textcolor[rgb]{0.00,0.00,0.00}{cannot} distinguish the users with similar angular direction in each cluster, larger $N_{\mathrm{T}}$ enables a higher gain in the communication links and interference links simultaneously. For {NFC}, by exploiting beamfocusing, the proposed {SLB-NF-NOMA} scheme can not only enhance the H-QoS user's signal strength at the focusing point, but also decrease the intra-cluster and inter-cluster interference to other users, even if the users lies in the similar angular direction.
	
	\subsection{{MLB-NF-NOMA} }  
	For {MLB-NF-NOMA}, the simulated geometry is set as follows: the AoD of the users in the four clusters are randomly distributed in the angle ranges $\theta _{1,k}\in \left[ -40^{\mathrm{o}},-30^{\mathrm{o}} \right] $, $\theta _{2,k}\in \left[ -15^{\mathrm{o}},-5^{\mathrm{o}} \right] $, $\theta _{3,k}\in \left[ 5^{\mathrm{o}},15^{\mathrm{o}} \right] 
	$ and $\theta _{4,k}\in \left[ 30^{\mathrm{o}},40^{\mathrm{o}} \right]  
	$, $m=1,2,3$. The radii of the users in the {MLB-NF-NOMA} system are set the same as the radii of the users in the {SLB-NF-NOMA} system and the AoD difference of the two users in each cluster are set to be $1^{\mathrm{o}}$. The minimum number of antennas allocated to each user is assumed to be $N_{\min}=0.2N$. Apart from NF-OMA scheme, we also consider the following benchmark schemes for comparison.
	\begin{enumerate}[(1)]  	
		\item '\textbf{Rand-MLB-NF-NOMA}' scheme, where the number of antennas allocated to the H-QoS user and L-QoS user are selected randomly.
		\item '\textbf{Fixed-MLB-NF-NOMA}' scheme, where $N_{\min}=0.2N$ antennas are allocated to user $\mathcal{U} \left( m, l\right) $ and $\left( N_{\mathrm{T}}-N_{\min} \right) 
		$ antennas are allocated to user $\mathcal{U} \left( m, h\right) $.
		\item '\textbf{MB-FF-NOMA}' scheme, where the analog beamformers are designed according to the multi-beam scheme proposed in~\cite{Wei2018MultiBeamNF} for FFC systems.
	\end{enumerate}
	
	For a fair comparison, the SVD-ZF digital beamformer and the proposed antenna allocation algorithm are adopted for all the benchmark schemes.
	\subsubsection{Average Sum Rate Versus $P_{\max}$}   
	\begin{figure}[t]
		\setlength{\belowcaptionskip}{-20pt}
		\centering
		\includegraphics[scale=0.6]{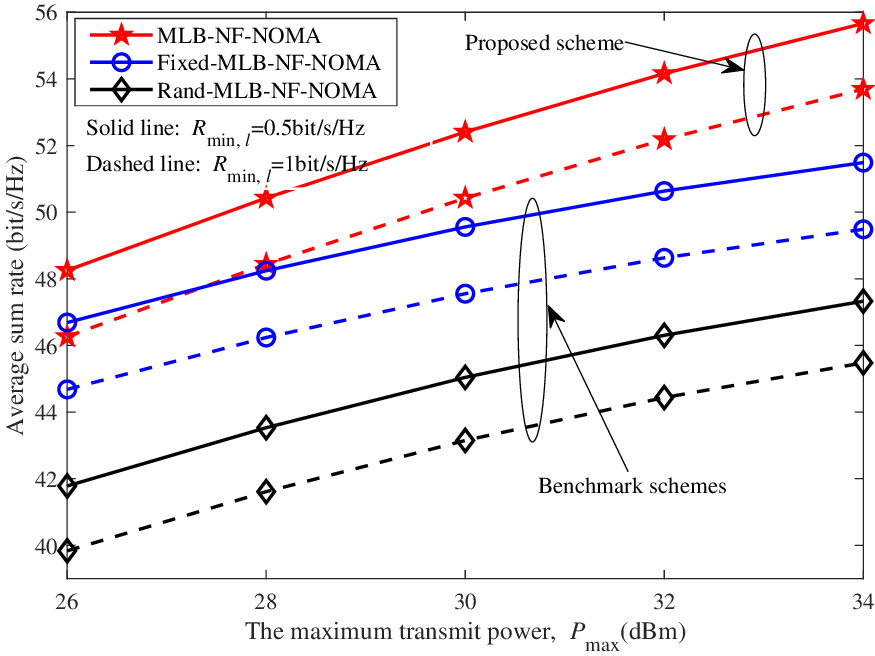}
		\caption{Average sum rate versus $P_{\max}$ for $N_{\mathrm{T}}=512$}% 
		\label{MB_sumR_vs_Pmax}
	\end{figure}
	
	To show the effectiveness of the proposed antenna allocation algorithm, we compare the proposed {MLB-NF-NOMA} scheme with the {Rand-MLB-NF-NOMA} and {Fixed-MLB-NF-NOMA} schemes. As it can been seen in Fig.~\ref{MB_sumR_vs_Pmax}, the proposed scheme achieves the highest sum rate performance. This reveals that the proposed many-to-one based antenna allocation algorithm can improve the system performance compared with the random and fixed antenna allocation schemes.  %he reason is that the MLB-NF-NOMA system can benefit from the antenna allocation strategy
	\subsubsection{Average Sum Rate Versus $N_{\mathrm{T}}$}  
	\begin{figure}[t]
		\setlength{\belowcaptionskip}{-20pt}
		\centering
		\includegraphics[scale=0.6]{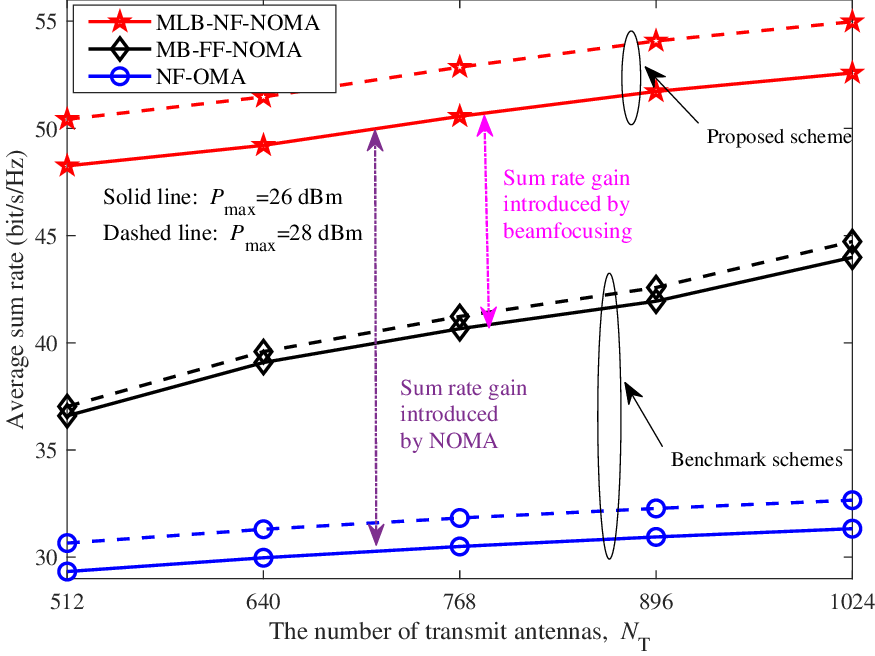}
		\caption{Average sum rate versus $N_{\mathrm{T}}$ for $R_{\min,l}=0.5$~bit/s/Hz} %
		\label{MB_sumR_vs_NT}
	\end{figure}
	
	In Fig.~\ref{MB_sumR_vs_NT}, we compare the average sum rate performance of various schemes versus $N_{\mathrm{T}}$. It is clear that when $N_{\mathrm{T}}$ increases, which means that more antennas are exploited to transmit the BS signal, higher possible beamfocusing and beamforming gain can be achieved by {MLB-NF-NOMA} and {MB-FF-NOMA} schemes, respectively. Moreover, the proposed MLB-NF-NOMA scheme outperforms the {MB-FF-NOMA} and NF-OMA schemes. 
	\subsubsection{Total Interference Versus $N_{\mathrm{T}}$} 
	\begin{figure}[t]
		\setlength{\belowcaptionskip}{-20pt}
		\centering
		\includegraphics[scale=0.6]{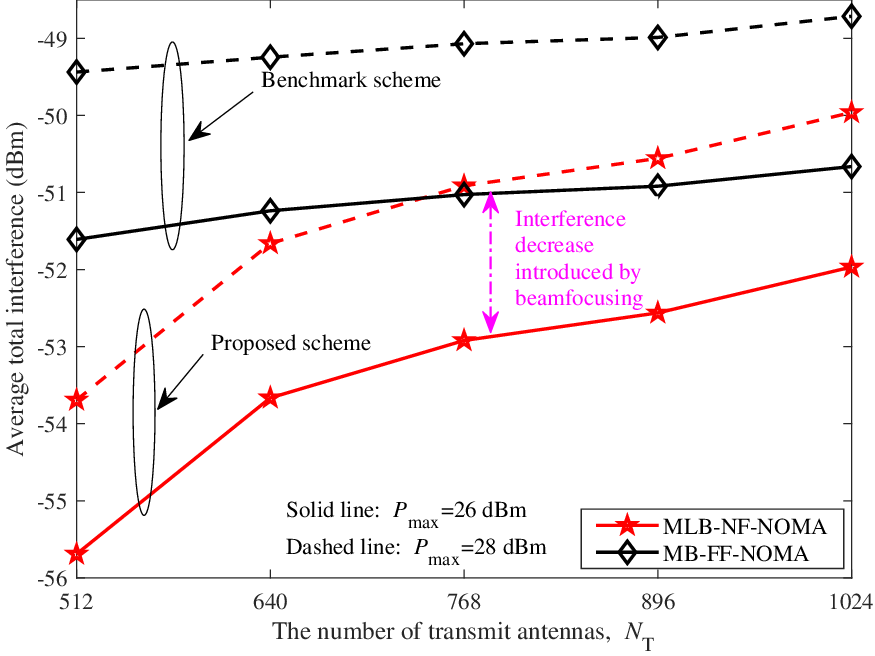}
		\caption{Average total interference versus $N_{\mathrm{T}}$ for $R_{\min,l}=0.5$~bit/s/Hz} %
		\label{MB_Interference_vs_NT}
	\end{figure}
	
	In Fig.~\ref{MB_Interference_vs_NT}, we further investigate the average total interference versus $N_{\mathrm{T}}$. Firstly, it is observed that the proposed {MLB-NF-NOMA} scheme can achieve lower interference compared with the {MB-FF-NOMA} scheme. This is indeed expected, employing NF-beamfocusing, {MLB-NF-NOMA} scheme can mitigates the total interference (including intra-cluster and inter-cluster interference) compared to the {MB-FF-NOMA} scheme. Secondly, average total interference of the two schemes increases with the increase of $N_{\mathrm{T}}$, which is quite different from the observation in Fig.~\ref{Interference_vs_NT}. The reason for this trend is \textcolor[rgb]{0.00,0.00,0.00}{explained} as follows. Compared to the {SLB-NF-NOMA} scheme, the beamwidth is increased after beam-splitting in {MLB-NF-NOMA} scheme, resulting in a higher intra- and inter-cluster interference. Since the intra- and inter-cluster interference can not be completely eliminated by the SVD-ZF based digital beamformer, increasing $N_{\mathrm{T}}$ can achieve a higher beam gain, while the total interference is also increased.
	\vspace{-0.5cm}
	\section{Conclusions}
	A novel NF-NOMA concept has been proposed, which can \textcolor[rgb]{0.00,0.00,0.00}{enhance the} flexibility of applying NOMA \textcolor[rgb]{0.00,0.00,0.00}{via exploiting the angular- and distance-domain DoFs in NFC.} Both {SLB-NF-NOMA} and {MLB-NF-NOMA} frameworks were proposed for \textcolor[rgb]{0.00,0.00,0.00}{hybrid beamforming transmitters. In SLB-NF-NOMA, the analog beamformer focusing on one specific location servers two NOMA users having the same angular direction in each cluster. In MLB-NF-NOMA, the analog beamformer focusing on two different locations servers two NOMA users having different angular directions in each cluster. For each framework, the H-QoS users' sum rate maximization problem was formulated, which was efficiently solved by the developed algorithms for obtaining the desired analog and digital beamformers and power allocation.} Numerical results
	confirmed that in contrast to the FFC schemes, the proposed NF-NOMA schemes can achieve better
	SE \textcolor[rgb]{0.00,0.00,0.00}{and provide a higher flexibility for applying NOMA transmission.} Moreover, our results revealed that NF-beamfocusing is an
	efficient means to mitigate the total interference {in comparison to FF-beamsteering}.
	\vspace{-0.2cm}
	\section*{Appendix~A: Proof of Lemma~\ref{inequialty_Gain}} \label{Proof_Lemma}
	\renewcommand{\theequation}{A.\arabic{equation}}
	\setcounter{equation}{0} 
	Recall the analog beamformer defined in~\eqref{single_beamfocusing} and the normalized antenna array gain obtained in~\eqref{AAgain_single}, we have that the maximum value of the antenna array gain for \textcolor[rgb]{0.00,0.00,0.00}{single-location-focused analog beamformer} is 1, \textcolor[rgb]{0.00,0.00,0.00}{i.e.,} $\left| \mathbf{b}^H\left( r_{m,h},\theta _{m,h} \right) \mathbf{w}_{m}^{\mathrm{SA}} \right|=1$. In addition, for \textcolor[rgb]{0.00,0.00,0.00}{multiple-location-focused analog beamformer} based on beam-splitting, we have:
	\begin{equation}\label{AAgain_multiple}
		\setlength{\abovedisplayskip}{5pt}
		\setlength{\belowdisplayskip}{5pt}
		\begin{split}
			\left| \mathbf{b}^H\left( r,\theta \right) \mathbf{w}_{m}^{\mathrm{MA}} \right| & =\frac{1}{N_{\mathrm{T}}}\left| \sum_{n=0}^{N_{m,h}-1}{e^{j\frac{2\pi}{\lambda}\left( r^{\left( n \right)}-r_{m,h}^{\left( n \right)} \right)}}+\sum_{n=N_{m,h}}^{N_{\mathrm{T}}-1}{e^{j\frac{2\pi}{\lambda}\left( r^{\left( n \right)}-r_{m,l}^{\left( n \right)} \right)}} \right|
			\\
			& =A^{\mathrm {MA}}_{\mathrm{gain}}\left( r,\theta ,\left\{ r_{m,k},\theta _{m,k},N_{m,k} \right\}  \right) , 
		\end{split}		 
	\end{equation}
	with the antenna array gain $A^{\mathrm {MA}}_{\mathrm{gain}}\left( r,\theta ,\left\{ r_{m,k},\theta _{m,k},N_{m,k} \right\} _{k\in \left\{ h,l \right\}} \right)$ defined as
	\begin{equation}\label{AAgain_multiple11}
		A_{\mathrm{gain}}^{\mathrm{MA}}\left( r,\theta ,\left\{ r_{m,k},\theta _{m,k},N_{m,k} \right\} _{k\in \left\{ h,l \right\}} \right) =\left\{ \begin{array}{c}
			\frac{1}{N_{\mathrm{T}}}\left| N_{m,h}+\widetilde{N}_{m,l} \right|,r=r_{m,h},\theta =\theta _{m,h},\\
			\frac{1}{N_{\mathrm{T}}}\left| N_{m,l}+\widetilde{N}_{m,h} \right|,r=r_{m,l},\theta =\theta _{m,l},\\
		\end{array} \right. 
	\end{equation}
	where $\widetilde{N}_{m,l}=\sum_{n=N_{m,h}}^{N_{\mathrm{T}}-1}{e^{j\frac{2\pi}{\lambda}\left( r_{m,h}^{\left( n \right)}-r_{m,l}^{\left( n \right)} \right)}}
	$ and $\widetilde{N}_{m,h}=\sum_{n=0}^{N_{m,h}-1}{e^{j\frac{2\pi}{\lambda}\left( r_{m,l}^{\left( n \right)}-r_{m,h}^{\left( n \right)} \right)}}
	$.
	
	It is easy to conclude that the function $\frac{1}{N_{\mathrm{T}}}\left| N_{m,h}+\widetilde{N}_{m,l} \right|$ satisfies the following inequalities		
	\begin{equation}\label{inequaltiy2}
		\frac{1}{N_{\mathrm{T}}}\left| N_{m,h}+\widetilde{N}_{m,l} \right|\overset{\left( a \right)}{\leqslant}\frac{1}{N_{\mathrm{T}}}\left( N_{m,h}+\left| \widetilde{N}_{m,l} \right| \right) \overset{\left( b \right)}{<}1,
	\end{equation}		
	where (a) comes from the inequality of complex modulus, \textcolor[rgb]{0.00,0.00,0.00}{and} (b) holds because the following inequality holds
	\begin{equation}\label{inequaltiy3}
		\left| \widetilde{N}_{m,l} \right|=\left| \sum_{n=N_{m,h}}^{N_{\mathrm{T}}-1}{e^{j\frac{2\pi}{\lambda}\left( r_{m,h}^{\left( n \right)}-r_{m,l}^{\left( n \right)} \right)}} \right|<N_{m,l},
	\end{equation}	
	where $r_{m,h}^{\left( n \right)}\ne r_{m,l}^{\left( n \right)}$ and $N_{m,h}+N_{m,l}=N_{\mathrm{T}}$.
	
	Similarly, for the function $\frac{1}{N_{\mathrm{T}}}\left| N_{m,l}+\widetilde{N}_{m,h} \right|$, we have	
	\begin{equation}\label{inequaltiy3}
		\frac{1}{N_{\mathrm{T}}}\left| N_{m,l}+\widetilde{N}_{m,h} \right|\leqslant \frac{1}{N_{\mathrm{T}}}\left( N_{m,l}+\left| \widetilde{N}_{m,h} \right| \right) <1.
	\end{equation}		
	
	Finally, Lemma~\ref{inequialty_Gain} is proved.
	
	\vspace{-0.2cm}
	\bibliographystyle{IEEEtran}
	\bibliography{zjkbib}
	
\end{document}